\documentclass[10pt,twocolumn,letterpaper]{article}
\usepackage[pagenumbers]{cvpr}

\usepackage{xspace}
\usepackage{tabularx}
\usepackage{stfloats}
\usepackage{multirow}
\usepackage{multicol}
\usepackage{makecell}
\usepackage{comment}
\usepackage{array}
\usepackage{booktabs}
\usepackage[table]{xcolor}
\usepackage{float}
\usepackage{balance}  % in preamble
\usepackage{placeins}
\usepackage{titletoc}

\definecolor{cvprblue}{rgb}{0.21,0.49,0.74}
\usepackage[pagebackref,breaklinks,colorlinks,allcolors=cvprblue]{hyperref}

\newcolumntype{P}[1]{>{\centering\arraybackslash}p{#1}}

\newcommand{\method}{Realiz3D\xspace}
\newcommand{\fullname}{Realiz3D\xspace}

\newcommand{\adapterfull}{Domain Shifter\xspace}
\newcommand{\adaptershort}{DS\xspace}
\newcommand{\figref}[1]{Fig.~\ref{#1}}

\makeatletter
\renewcommand{\paragraph}{%
  \@startsection{paragraph}{4}%
  {\z@}{-0.5em}{-0.5em}%
  {\normalfont\normalsize\bfseries}%
}
\makeatother

\def\paperID{}
\def\confName{CVPR}
\def\confYear{2026}

\title{\fullname: 3D Generation Made Photorealistic via Domain-Aware Learning}

\author{\normalsize Ido Sobol$^{1,2}$ \quad
Kihyuk Sohn$^2$ \quad Yoav Blum$^2$ \quad Egor Zakharov$^2$ \quad Max Bluvstein$^2$ \quad Andrea Vedaldi$^2$ \quad Or Litany$^{1}$
\vspace{0.2cm} \\
{\normalsize $^1$ Technion \quad $^2$ Meta AI}
\\
}

\begin{document}

\twocolumn[{%
\renewcommand\twocolumn[1][]{#1}%
\maketitle
\begin{center}
\includegraphics[width=0.95\textwidth]{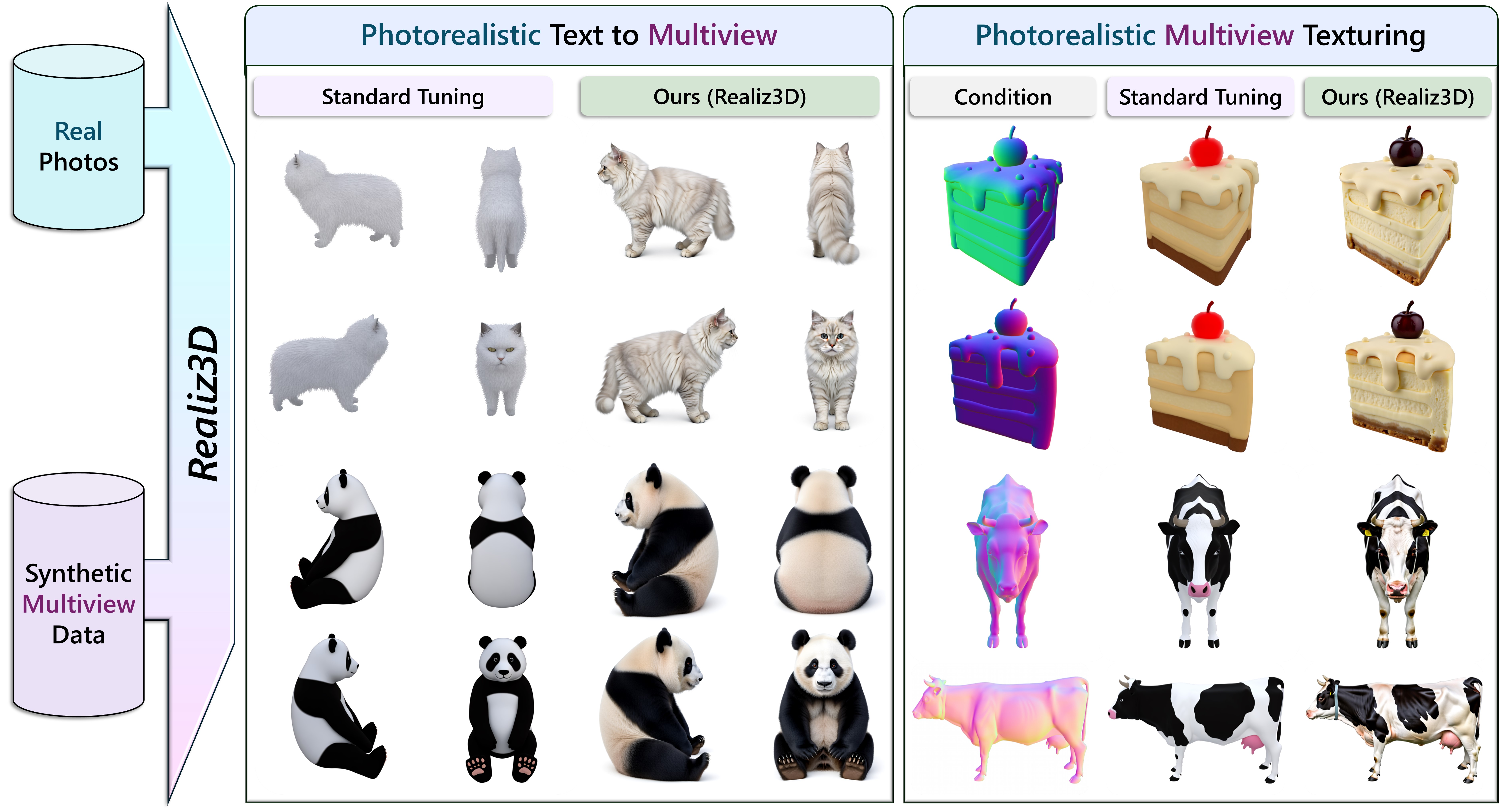}
\captionof{figure}{\emph{\method} is a framework that leverages both real and synthetic data to train diffusion models that generate photorealistic images while faithfully adhering to input conditions and maintaining 3D consistency.  
Shown are two representative applications: text-to-multiview generation (left) and multiview texturing (right).  
Compared to standard fine-tuning on mixed real and synthetic data, \method produces noticeably more realistic results while preserving geometric fidelity across views.  
The prompts shown are (left) ``a fluffy cat'', ``a panda bear, sitting''; (right) ``a slice of creamy cheesecake with a cherry on top'', ``a cow with black and white patterns''. Best viewed zoomed in.}%
\label{fig:teaser}
\end{center}
}]

\begin{abstract}
We often aim to generate images that are both photorealistic and \emph{3D-consistent}, adhering to precise geometry, material, and viewpoint controls.
Typically, this is achieved by fine-tuning an image generator, pre-trained on billions of real images, using renders of synthetic 3D assets, where annotations for control signals are available.
While this approach can learn the desired controls, it often compromises the realism of the images due to domain gap between photographs and renders. 
We observe that this issue largely arises from the model learning an unintended association between the presence of control signals and the synthetic appearance of the images.
To address this, we introduce \emph{\fullname}, a lightweight framework for training diffusion models, that decouples controls and visual domain.
The key idea is to explicitly learn visual domain, real or synthetic, separately from other control signals by introducing a co-variate that, fed into small residual adapters, shifts the domain. Then, the generator can be trained to gain controllability, without fitting to specific visual domain.
In this way, the model can be guided to produce realistic images even when controls are applied.
We enhance control transferability to the real domain by leveraging insights about roles of different layers and denoising steps in diffusion-based generators, informing new training and inference strategies that further mitigate the gap.
We demonstrate the advantages of \fullname in tasks as text-to-multiview generation and texturing from 3D inputs, producing outputs that are 3D-consistent and photorealistic.
\end{abstract}
\vspace{-2em}
\section{Introduction}%
\label{sec:introduction}

While diffusion-based image generators have made significant progress in recent years, it remains challenging to equip them with \emph{precise 3D controls} to ensure that images conform to prescribed geometry, material, and viewpoint specifications.
The latter is important, or even crucial, in many applications.
For example, image generators are often used to guide 3D content generation~\cite{mvdream, liu2023zero, bensadoun2024meta, tang2024lgm, gslrm2024} in an attempt to sidestep the lack of large-scale 3D datasets to train such models directly.
There is no lack of data to train image generators; however, their application to 3D generation requires the ability to control geometry and viewpoint, which is not natively supported by most image generators.

The main challenge in training image generators with 3D controls is that real images lack 3D-like annotations such as geometry, materials, and cameras.
Thus, the usual strategy is to pre-train the model on billions of real images and then fine-tune it on a relatively small number of renders of synthetic 3D assets, for which 3D annotations can be easily obtained.
However, such renders are far from photorealistic, resulting in a severe domain gap compared to real images.
Ultimately, this leads to an undesirable trade-off between \emph{realism} learned from real images and \emph{controllability} learned from synthetic 3D data.
We study this problem and identify a key cause for the degradation in realism: when fine-tuning on synthetic data, the model tends to associate the presence of 3D controls with the synthetic look of the corresponding images.
In other words, the control signals \emph{leak domain identity}, so that, when controls are given at inference, the model \emph{also} makes the image look synthetic.

To address this problem, we introduce \fullname, a lightweight framework for fine-tuning image generators with 3D control, while preserving photorealism.
We do so by decoupling domain identity from control signals.
% Before learning the desired 3D controls, we learn a separate control for the domain itself.
% The latter is a binary indicator:
% for real images, it tells the model to operate in `real mode', while for synthetic images, it tells the model to operate in `synthetic mode'.
% This information is incorporated by adding a small residual adapter that `shifts' the visual domain according to the value of the domain control.
In the first stage, before learning the desired 3D controls, we learn a separate binary control for visual domains, indicating the model to operate in real or synthetic mode.
Domain signal is integrated through \adapterfull{s}, lightweight residuals that shift the generation toward the desired visual domain.

After the first stage, we introduce the desired 3D control signals (e.g., one or multiple viewpoints, normal maps, or other cues), using the synthetic samples for supervision.
Although 3D annotations are available only for synthetic data, the \adapterfull have already learned to disentangle visual domains as a co-variate introduced before, reducing domain leakage.
At inference, we can operate in the ``real mode'', while providing the model with 3D control signals, yielding results that are both realistic and controllable.

To achieve effective transfer of control to the real domain, while maintaining realism, we leverage insights into the roles of different network layers and denoising steps in diffusion-based image generators.
As observed before~\cite{tumanyan2023plug,luo2023diffusion, jiang2025no}, early network layers and denoising steps predominantly determine the structure of the generated image, whereas later layers and steps determine its detailed appearance.
\fullname exploits this by allowing synthetic data to influence early layers and denoising steps more strongly, while real data has a stronger effect on later ones.

Together, these stages encourage domain-agnostic behavior, enabling effective transfer of control to real domain.

To summarize, our contributions are threefold:
\begin{enumerate}
\item A flexible and general recipe for tuning diffusion models on controllable yet domain-shifted datasets, while maintaining the realistic prior of the base model.

\item A new domain-shifting adapter design that separates domain identity from control signals and prevents domain leakage during fine-tuning.

\item A layer-aware training and sampling strategy that progressively unifies feature spaces, enabling realistic and controllable generation for tasks such as text-to-multiview and texturing from 3D inputs.
\end{enumerate}

\vspace{-0.5em}
\section{Related Work}%
\label{sec:related-work}

\paragraph{Control in Image and 3D Generation.}

While the quality of image generators has improved dramatically in recent years, control is equally important in applications.
Many have thus sought to augment image generators with control signals like depth or normal maps, semantic masks, camera viewpoints, or human poses, to enable conditional generation.
These methods typically inject control into pre-trained models and fine-tune to achieve controllability~\cite{zhang2023adding,hu2022lora,mou2024t2i,li2023gligen}.
Learning 3D controls (e.g., depth maps, normals, or multiple viewpoints)~\cite{mvdream,shi2023zero123++,long2024wonder3d,xu2025flexgen,lin2025kiss3dgen} requires data annotated with this information, which is difficult to obtain in the real world.
Hence, authors often use synthetic data, for example by rendering assets in large-scale 3D model collections like Objaverse~\cite{deitke2023objaverse,deitke2024objaverse}.
However, fine-tuning a model on synthetic data can affect realism, `forgetting' the look of real images.
Various approaches were proposed to mitigate forgetting, including LoRA layers~\cite{hu2021lora}, adapters and ControlNet modules~\cite{zhang2023adding}, or simply by `replaying' real data while fine-tuning on synthetic data~\cite{mvdream}.

\paragraph{Training Adapters To Mitigate domain Gaps.}

\noindent \textit{Wonder3D}~\cite{long2024wonder3d} jointly generates multiview RGB images and corresponding normal maps by introducing a domain switcher that modifies the model’s existing conditioning mechanism. A 1D domain vector is concatenated with the timestep embedding and learned jointly with the model. However, this method does not explicitly enforce consistency between the generated image and its normal map, relying instead on synthetic paired data and cross-domain attention.
% , between an image and its normal map.
Similarly, we learn domain embeddings to guide the model towards one of multiple domains.
Unlike~\cite{long2024wonder3d}, we neither modify the existing conditioning mechanism nor rely on paired data. Moreover, since both realistic and synthetic domains are well represented in T2I models, jointly training the adapters with the model may cause it to collapse into two modes (controllable and synthetic, vs realistic and uncontrollable).
\textit{AnimateDiff}~\cite{guo2023animatediff} adapts T2I models for video generation using video data, which is often lower quality than image datasets. They train domain adapters, implemented as LoRA layers, to first fit the noisy domain using video frames, then freeze the adapters while training the model on videos. The adapters are removed at inference. We also use a multi-stage training procedure, but fitting an adapter to the synthetic domain is ineffective, as the “undesired” domain is already encoded in the base model’s weights.
\textit{Still-Moving}~\cite{chefer2024still} trains temporal attention blocks to adapt a T2I model for video generation, and then reuses them in a customized T2I model. To align the temporal blocks’ outputs with the model’s distribution, they introduce Spatial Adapters implemented as linear projections.
\vspace{-0.5em}
\section{Diffusion Models and Domain Gaps}%
\label{sec:preliminaries}

We consider image generators based on denoising diffusion~\cite{song2020denoising} and summarize key findings from the literature.

\paragraph{Timesteps and Domain Gaps.}
In denoising diffusion models, sampling evolves through timesteps $t = T, T-1, \dots, 0$. Starting from $X_T \sim \mathcal{N}(0, I)$, a neural network $\Phi$ iteratively denoises $X_t$ at each timestep to generate a clean data sample $X_0$.
Previous studies~\cite{meng2021sdedit, yang2023diffusion, yi2024towards, sobol2024zero, peng2024lesson} show that early timesteps ($t \,{\approx}\, T$) primarily establish low-frequency structure of generated samples, while later timesteps ($t \,{\approx}\, 0$) determine high-frequency details.
Formally, consider two marginal distributions
$
q(X^\text{real}_t)
$
and
$
q(X^\text{syn}_t)
$
obtained by noising real and synthetic distributions
$
p(X^\text{real})
$
and
$
p(X^\text{syn})
$.
In the limit of $t = T$, both distributions converge to the same Gaussian distribution, thus are equal.
SDEdit~\cite{meng2021sdedit} adds noise to a non-realistic image, and denoises with a pre-trained diffusion model, to produce a realistic image that preserves the structure.
\cite{daras2023ambient} shows that noisy data can be used for training at early timesteps.

\paragraph{Layers and Domain Gaps.}

The level of details across generation in  diffusion models is not only linked to the timestep $t$, but also to the layers of the underlying denoising neural network.
\cite{tumanyan2023plug, luo2023diffusion, jiang2025no, stern2025appreciate} have studied the feature maps computed by different layers of UNet-based diffusion models.
UNet~\cite{ronneberger2015u} is a hierarchical encoder-decoder architecture with skip connections.
The encoder extracts progressively coarser structures, while the decoder upsamples and combines features to integrate both coarse and fine-grained information in the output.
\cite{tumanyan2023plug} shows that low-resolution UNet features capture rough 2D shapes and low-frequency patterns, while high-resolution features encode textures and fine details. \cite{luo2023diffusion} demonstrates that feature maps capture progressively finer details as denoising advances.
Others have noted similar patterns in denoising diffusion transformers~\cite{jiang2025no}, and in vision transformers in general~\cite{ghiasi2022vision, amir2021deep}.

In this work, we leverage these insights by enforcing 3D controls in earlier layers of a \textit{diffusion transformer}~\cite{peebles2023scalable}, while allowing deeper layers to maintain realism.

\vspace{-0.5em}
\section{Method}%
\label{sec:method}

\paragraph{Problem Formulation.}
%new ver
We aim to train a controllable diffusion model capable of generating \(V \ge 1\) photorealistic and 3D-consistent views 
\( \{ x_{\text{real}}^{v} \}_{v=1}^{V}\), 
conditioned on one or more spatial control signals \(c\), such as per-view normal or depth maps. 
Formally, the model learns the conditional distribution
\(
q_{\theta}\big(\{x_{\text{real}}^{v}\}_{v=1}^{V} \mid c\big)
\)
that generates realistic and controlled samples, geometrically consistent across views.

To achieve this, we assume access to a text conditioned image generator
\(
q_{\psi} \big( x_{\text{real}})
\)
pre-trained on real images and two complementary data sources: 
a \textit{synthetic} dataset \( \{ \{x_{\text{syn}}^v\}_{v=1}^{V}, c \} \), rendered from 3D assets that provide accurate supervision for the control signal \(c\) (e.g., camera pose, normals); 
and a \textit{real} dataset \( \{ x_{\text{real}}, \varnothing \} \), composed of diverse natural images with null control signal $c = \varnothing$.

\paragraph{Method Overview.}
We train an image generator for multiview synthesis by extending its single-view output to a grid representation~\cite{shi2023zero123++, bensadoun2024meta}, where multiple views are spatially tiled, and self-attention operates between all views. For real data, we form grids of arbitrary images and restrict attention to operate within each view (\textit{single-image mode})~\cite{mvdream}.

The na\"{\i}ve approach of fine-tuning $q_\theta$ on synthetic data 
$\{\{x_{\text{syn}}^v\}_{v=1}^{V}, c\}$ 
alone can successfully learn the control signal $c$, but may \emph{catastrophically forget} the appearance of realistic images due to overfitting to synthetic images. 
Mixed-domain training, which mixes the real data without control signal with the synthetic data, mitigates, but does not fully resolve the forgetting issue, as shown in \cref{eval:texturing_quantitative}.
We hypothesize that, since only the synthetic samples carry non-null control, the model implicitly associates the very presence of $c\neq\varnothing$ with the synthetic domain, causing leakage of synthetic appearance whenever control is applied.

\begin{figure*}[ht]
    \centering
    \centerline{
        \includegraphics[width=0.9\linewidth]{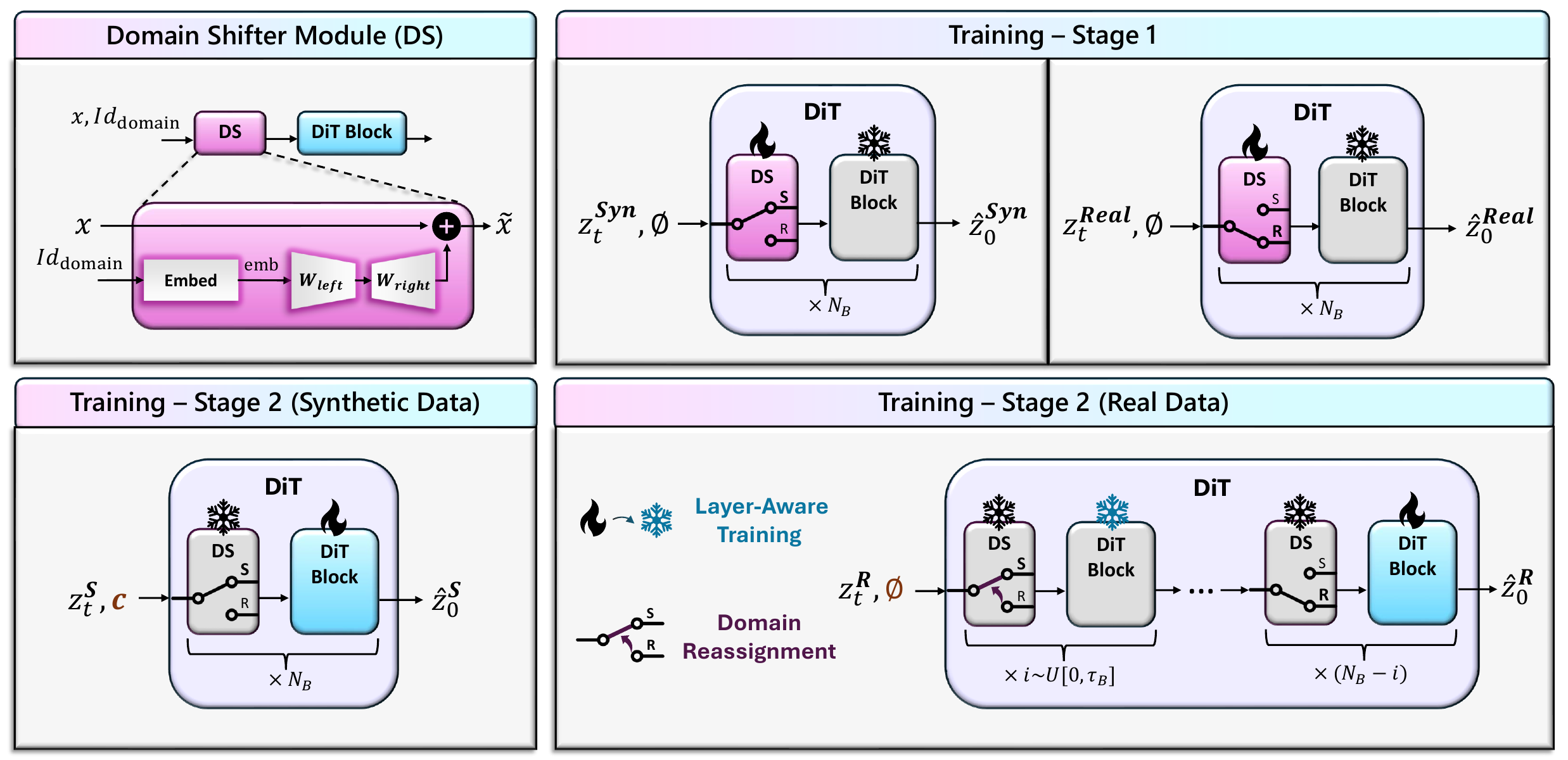}
    }
    \vspace{-0.5em}
    \caption{\textbf{Method overview.} \method introduces \adapterfull{s}, lightweight residual adapters that learn visual domain identity (real vs. synthetic) independently of control signals, enabling the model to learn controllability without compromising realism.
\textbf{(top left)} A \adapterfull encodes domain identity as a low-rank residual added to latent features (Sec.~\ref{subsec:dom}).
\textbf{(top right)} Stage~1: \adapterfull{s} are trained with mixed real and synthetic data under null control while the base model is frozen, learning domain separation (Sec.~\ref{subsec:dom}).
\textbf{(bottom)} Stage~2: The diffusion model is fine-tuned for controllable generation using both domains (Sec.~\ref{subsec:layer_binding}).
\textbf{(bottom left)} Synthetic samples teach controllability under the synthetic mode.
\textbf{(bottom right)} Real samples (without control) are used for \textit{Representation Binding}—combining (1) \textit{Layer-Aware Training}, which freezes some early layers to preserve structure while updating later, appearance-related ones, and (2) \textit{Domain Reassignment}, which occasionally reuses the synthetic mode in early layers to transfer control to the real domain. Throughout all stages of training, we rely solely on the standard diffusion loss} % Throughout all stages of training, we rely solely on the standard diffusion loss.
    \label{fig:scheme}
\end{figure*}

To address this issue, we explicitly separate domain identity from the control signal by introducing a co-variate 
$e_{\text{domain}} \in \{e_{\text{real}}, e_{\text{syn}}\}$, 
injected into the model via our \textit{\adapterfull{s}} (\cref{fig:scheme}). 
In \textit{stage~1}, we freeze the diffusion backbone and train only the \adapterfull{s} to distinguish between real and synthetic data under null control, learning
\(
q_\theta(x \mid e_{\text{domain}}, \varnothing),
\)
so that the model internalizes the notion of domain \emph{independently of control}. 
In \textit{stage~2}, we introduce control conditioning (available only for synthetic data) and fine-tune the shared backbone to follow it, modeling
\(
q_\theta(\{x^v\}_{v=1}^V \mid e_{\text{domain}}, c).
\)
Utilizing the \adapterfull{s}, we propose a \textit{Representation Binding} strategy that ensures that controllability learned on synthetic data transfers effectively to the realistic domain.

\noindent Throughout the training, \textit{we rely solely on the standard diffusion loss}. The diffusion objective is used to train our \adapterfull{s} in Stage 1 and the DiT backbone in Stage 2.

\noindent \textit{At inference time}, setting $e_\text{domain}=e_\text{real}$ with $c\neq\varnothing$ enables the model to generate realistic yet controllable images.

\vspace{-0.2em}
\subsection{Decoupling Domain from Control with \adapterfull{s} (Stage~1)}%
\label{subsec:dom}

Given an image generator based on denoising diffusion, we denote by $X \in \mathbb{R}^d$ the latent representation entering a diffusion block. A \emph{\adapterfull} module $\mathcal{D}$ consists of two learnable domain embeddings, $e_{\text{syn}}, e_{\text{real}} \in \mathbb{R}^d$, and a shared low-rank transformation that maps these embeddings into the model's latent space by applying a domain-specific residual adapter (See top row in~\cref{fig:scheme}):
\[
\tilde{X} = X + \mathcal{D}(\text{domain}) \;=\; X + W_{\text{left}} W_{\text{right}}\, e_{\text{domain}},
\]
where $W_{\text{left}}\in\mathbb{R}^{d\times r}$ and $W_{\text{right}}\in\mathbb{R}^{r\times d}$ define a rank-$r$ mapping with $r \ll d$. 

\noindent The embedding is added to all tokens within the block, acting as a low-rank bias that modulates activations according to domain identity. 
Analogous to LoRA-style adapters, this low-rank residual provides sufficient capacity to traverse nearby modes in latent space~\cite{rebuffi17learning} while maintaining stability and efficiency.
In stage 1 (\cref{fig:scheme}, top right) we freeze the diffusion backbone and optimize only \adapterfull{}s using both real and synthetic images with null control $c=\varnothing$. We operate in single-image mode, restricting attention to operate within each view only.
Since both domains already reside within the pre-trained model’s feature space, these lightweight low-rank residuals suffice to capture domain identity explicitly (further discussion is in the Appendix). Then, the model cleanly separates visual domains from control signals, laying the foundation for controllable cross-domain generation in stage 2.

\vspace{-0.2em}
\subsection{Fine-tuning with Representation Binding (Stage~2)}%
\label{subsec:layer_binding}

\paragraph{Backbone Fine-Tuning.}
To gain controllability without hindering realism, in stage~2 we propose a strategy for fine-tuning the diffusion backbone while keeping the \adapterfull{}s frozen.
A straightforward approach is to fine-tune the backbone with synthetic data only, while switching the \adapterfull{s} to synthetic mode, relying on the shared backbone to transfer controllability to real images.

At generation, however, we observe that when switching \adapterfull{s} to real mode, control signal is not always respected, and generated samples may still appear synthetic.
We attribute this to:
(1) \emph{Forgetting of realism}: the backbone drifts toward synthetic statistics since fine-tuning updates are applied only to synthetic data (see ablation 2 in Tab.~\ref{eval:texturing_ablation}); and
(2) \emph{Partial control transfer}: control transferability is merely emergent. Without access to samples with both $e_\text{domain}=e_\text{real}$ and $c\neq\varnothing$, the shared model lacks experience applying control under real-domain conditions again fitting to the synthetic distribution. 

Both factors highlight the need for domain-agnostic behavior in the shared model. To that end, we reintroduce real data for training during Stage 2, and propose a strategy that leverages the model’s internal feature hierarchy to enable robust transfer of control to the real domain.

\paragraph{Bridging Unpaired Domains through Feature Space.}
To address both challenges, we observe that early diffusion layers tend to be domain-agnostic: capturing coarse structure and low-frequency content, shared across real and synthetic images (\cref{sec:preliminaries}).
Later layers, in contrast, refine high-frequency appearance, where domain gap is more pronounced.
By leveraging early layers as a bridge, we can explicitly bind both domains in feature space, promoting transfer of controllability from synthetic to real data while preserving visual fidelity.
Building on this, we introduce two complementary strategies, shown in~\cref{fig:scheme} (bottom), that operationalize this principle: one preserves realism, and another enhances control transferability to real domain.

\noindent\textbf{(1)~Preserving Realism with Layer-Aware Training.}
To prevent forgetting of realism, we incorporate real samples into training, inspired by~\cite{mvdream}.
However, since real images lack explicit control supervision, naïvely training on them could interfere with the model’s ability to respect the control signal learned from synthetic data.
Guided by our observation that early layers are largely domain-agnostic and structure related, we update the model with real samples only in the later diffusion blocks, those primarily responsible for appearance refinement, while keeping early blocks frozen.
This ensures that training on real data does not disturb the control-related representations formed in early layers.

When processing real samples, \adapterfull{s} operate in real mode, allowing the model to maintain realistic appearance statistics without altering the shared structural pathway.
Concretely, during each real-data training iteration, we freeze DiT blocks $B \in [0, B_i]$, where $i$ is an integer block index randomly drawn from $[0, \tau_B]$ (see~\cref{fig:scheme}, bottom right).
This stochastic layer-freezing regularizes early representations, without requiring a fixed cutoff.

\noindent\textbf{(2)~Enhancing Control Transferability via Domain Reassignment.}
To further promote control transfer, we introduce \textit{Domain Reassignment}.
With probability $p_B$, we reassign early DiT blocks ($B\in[0,B_j]$, $j$ is an integer, sampled from $[0, \tau_B]$) to operate in synthetic mode even when processing real samples; that is, we substitute $e_\text{domain}\leftarrow e_\text{syn}$ in the corresponding \adapterfull{}s (\cref{fig:scheme}, bottom right).
This asymmetric design integrates real samples into the synthetic feature space, rather than the other way around, since the synthetic domain is the one endowed with explicit control supervision.
Consequently, early layers learn shared structural representations that carry controllability, while later layers remain anchored to real-domain appearance.

\noindent These components, shown in~\cref{fig:scheme}, form our \textit{Representation Binding} strategy: a soft feature-space alignment, preserving realism while encouraging control transfer.

\vspace{-0.2em}
\subsection{Inference-time Domain Shifting}
\label{subsec:sampling}
At inference time, thanks to the control transfer established during fine-tuning, one can simply switch the domain adapter to the \emph{real} mode ($e_\text{domain}=e_\text{real}$) and provide a control condition $c\neq\varnothing$.
The model then generates outputs that are both realistic in appearance and faithful to the specified control.
\textit{Yet, we can do even better.}
While this setup already enables controllable generation in real domain, we find that the control signal can be further strengthened without sacrificing realism.
As noted in \cref{subsec:layer_binding}, samples generated with $e_\text{domain}=e_\text{syn}$ tend to follow the control $c$ more faithfully, as the synthetic domain is \textit{directly} supervised for control.

At inference, we adopt a partial, non-stochastic \emph{domain reassignment}: pre-defined selected early layers and timesteps are set to synthetic mode ($e_\text{domain}=e_\text{syn}$), while later layers and timesteps remain in real mode. The configuration is tuned once, and not per example.
As early layers primarily capture coarse, domain-agnostic structure, this hybrid configuration allows users to rebalance realism and controllability at test time without additional training. Further details and tuning strategy appear in the Appendix.

\vspace{-0.5em}
\section{Experiments}%
\label{sec:experiments}

We demonstrate the effectiveness of \method on Multiview Texturing and Text-to-Multiview Generation.

\paragraph{Datasets.}

\noindent\textit{Synthetic Data:}
We use an internal dataset of $120$K synthetic 3D assets, with their textual descriptions.
Each asset is rendered from $V=4$ viewpoints, with normal and position maps.
\noindent\textit{Real Data:}
While we could use the training data of the base model, we simply use images generated by the base model itself. We use the textual descriptions from the synthetic dataset as prompts, ensuring fairness, and generate $V$ photorealistic images per prompt with white background.
The synthetic and real datasets are matched in size.
\noindent\textit{Evaluation Data:}
To evaluate our method, we use 40 3D objects, from Sketchfab, used and reported in~\cite{bensadoun2024meta}, along with prompts describing the original object and texture.
We create synthetic data (via rendering) and realistic data (by generating realistic images with the T2I model and text prompts) for the evaluation objects.

\paragraph{Implementation Details.}
We leverage an internal pre-trained text-to-image diffusion transformer, whose architecture follows~\cite{peebles2023scalable} and performance is comparable to other diffusion transformers such as Stable Diffusion 3~\cite{esser2024scaling}.
For all tasks, we fine-tune the model to generate a $2\times2$ image grid of $V=4$ orthogonal views, each at $512\times512$ resolution. Full implementation details are in the Appendix.

\paragraph{Baselines.}
We compare \method to three categories of adaptation techniques, trained under identical settings using the same base model and data. Then, we evaluate \method against pretrained models, producing multiview images from text. While we do not have access to their training data, they serve as reference points for assessing \method relative to state-of-the-art performance.

\noindent \emph{Full Fine-Tuning:}
All base model's parameters are trained.
(1) Using synthetic samples only (Syn only).
(2) Using synthetic and realistic samples~\cite{mvdream}, evenly (Syn + Real).

\noindent\emph{Lightweight Fine-Tuning:}
Only a small set of parameters is trained to retain the base model’s prior, using synthetic data only.
(3) Training LoRA layers~\cite{hu2021lora}, added to each transform in all attention layers.
% We test two common ranks: 32 and 128.
(4) Training linear adapter layers, added before each DiT block, implemented as the product of two low-rank matrices, inspired by~\cite{mou2024t2i}.

\noindent\emph{Adapter-based Methods}:
We evaluate methods, discussed in Sec.~\ref{sec:related-work}, that propose adapters to bridge domain gaps, and adjust them as necessary to tackle synthetic-to-real adaptation.
(5) Domain Adapters~\cite{guo2023animatediff}.
(6) Spatial Adapter~\cite{chefer2024still}, applied similarly to~\cite{guo2023animatediff} using linear adapter layers.
For (3)-(6) we test two common ranks: 32 and 128, and show results with rank 32.
(7) Two variants of Domain Switcher~\cite{long2024wonder3d}: one with the switcher trained jointly with the model~\cite{long2024wonder3d}, and another trained separately in a two-stage manner.

\noindent\emph{Training-Free Methods}: We evaluate SDEdit~\cite{meng2021sdedit} for multiview applications, using $t=500$ (See Appendix).

\noindent\emph{Pretrained (Text-to-Multiview).} We evaluate the realism of TRELLIS~\cite{Xiang2024Structured3L}, a recent 3D-native model.

% \vspace{-0.5em}
\subsection{Multiview Texturing}

\paragraph{Task Definition.}

Given a set of $V\geq1$ aligned normal maps $\{c_{normal}\}_0^V$ and position maps $\{c_{position}\}_0^V$, our goal is to generate 3D-consistent images $\{x_{RGB}\}_0^V$ matching the given geometry, thus learning the conditional distribution $q(\{x_{RGB}\}_0^V|\{c_{normal}\}_0^V, \{c_{position}\}_0^V)$.

\noindent \textbf{Implementation Details}:
Following previous works, normal and position maps are encoded using the model's VAE, and are channel-concatenated to the noisy latent.

\noindent \textbf{Metrics.}
\noindent\textit{3D Consistency:}
Following prior works, we backproject the generated images onto their mesh and project the same views, reporting PSNR, SSIM~\cite{ssim_2004} and LPIPS~\cite{lpips_cvpr_2018} between generated and re-projected views.

\noindent\textit{Prior Preservation:}
We compare our generated views to a set of realistic images generated by the base T2I model using the same prompts, and report $FID_B$ and $KID_B$.

\noindent\textit{Real-World Realism:}
Prior preservation metrics assess realism but rely on synthesized data. We compare our generated views to real-world images from ImageNet~\cite{deng2009imagenet}, selected from categories best fitting the evaluation objects (further details in the Appendix). We report $FID_I$ and $KID_I$.

\noindent\textit{Text-Image Alignment:}
We report CLIP score. The prompts describe the original texture, and do not require realism.

\noindent \textbf{Quantitative Evaluation.}
Main results in \cref{eval:texturing_quantitative} demonstrate that control and realism typically trade off. \method uniquely achieves strong performance in both.
Still, while \method achieves strong results with notably improved realism, its adherence to control remains slightly lower than fully synthetic baselines.
We attribute this to: (1) 3D consistency metrics are sensitive to fine-grained details---typically absent in synthetic data---that require precise pixel-level consistency; (2) Lack of realism can occasionally manifest in unrealistic geometry, causing \method to deviate from the signal in favor of photorealism; (3) Inconsistent appearance may result from the base model’s lighting bias.

\begin{figure*}
\centering
\centerline{
    \includegraphics[width=0.9\linewidth]{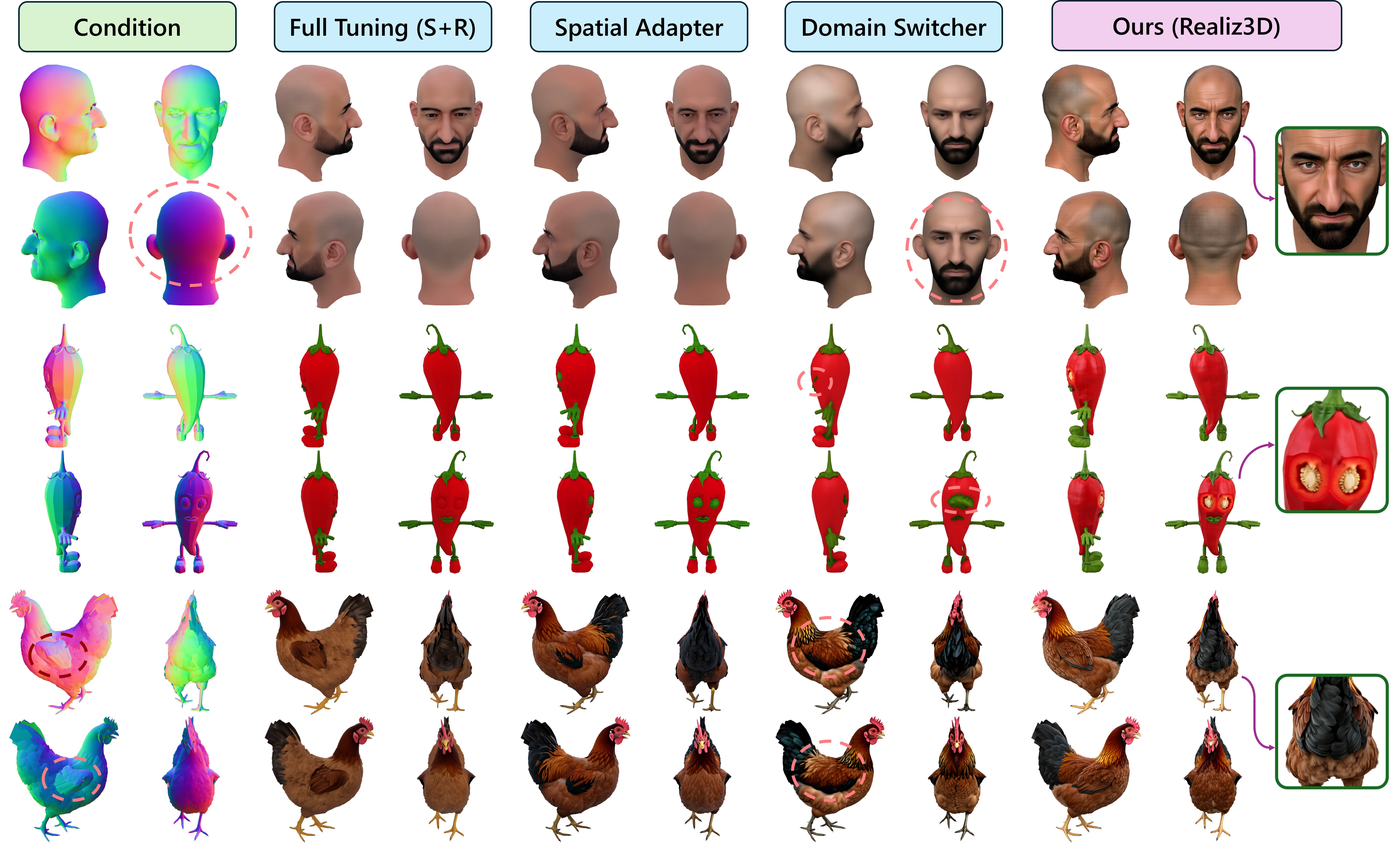}
}
\vspace{-0.5em}
\caption{\textbf{Multiview Texturing: Qualitative Results.} Prompts: "A man with dark, short beard", "A standing red pepper", "A brown barn chicken", respectively. All prompts are appended with “highly photorealistic and detailed”. Red, dashed circles highlight inconsistent regions (either with the geometry or with other views). \method achieves significant improvements in photorealism while remaining 3D-consistent and faithful to the geometric conditions. Only normal maps are shown due to limited space. Best viewed zoomed in.}%
\label{fig:texturing_baselines}
\end{figure*}

%%%%%%%%%%%%%%%
\noindent \textbf{Qualitative Evaluation.}
We present visual examples of generated textures in \cref{fig:teaser} and \cref{fig:texturing_baselines}. Only the best-performing baselines are shown in the main paper. Full and additional results appear in the Appendix.
To showcase 3D consistency, meshes textured with our generated images appear on our project page.

\noindent \textbf{Ablation Study.}
We firstly verify the importance of our two-stage training, and of using real data at stage 2. Additionally, to assess each component’s contribution, we evaluate \method by gradually adding them, showing that each enhances controllability with minimal compromise to realism, and that all together yield the best overall performance. Full results of the ablation study are reported in~\cref{eval:texturing_ablation}.

\begin{table*}
\centering
\caption{\textbf{Multiview Texturing: Quantitative Results.} \method achieves significantly improved realism while maintaining 3D consistency comparable to the synthetic-only baseline (top row).}
\vspace{-0.5em}
\label{eval:texturing_quantitative}
\resizebox{\textwidth}{!}{%
\begin{tabular}{c c c c c c c c c c}
\toprule
\multirow{2}{*}{} &
\multirow{2}{*}{Method} &
\multicolumn{3}{c}{\cellcolor{cyan!20} 3D Consistency} & 
\multicolumn{2}{c}{\cellcolor{magenta!20} Prior Preservation} & 
\multicolumn{2}{c}{\cellcolor{orange!20} Real-World Realism} &
\multicolumn{1}{c}{\cellcolor{yellow!20} Text Align} \\
\cmidrule(lr){3-5} \cmidrule(lr){6-7} \cmidrule(lr){8-9} \cmidrule(lr){10-10}
 & & PSNR $\uparrow$ & SSIM $\uparrow$ & LPIPS $\downarrow$ & FID$_{B}$ $\downarrow$ & KID$_{B}$ $\downarrow$ &
 FID$_{I}$ $\downarrow$ & KID$_{I}$ $\downarrow$ & CLIP $\uparrow$ \\
\midrule
\multirow{2}{*}{\makecell{Full\\Tuning}} 
  & Syn Only & \textbf{25.76} & \textbf{0.9269} & \textbf{0.0831} & 168.21 & 0.0240 & 218.29 & 0.0431 & 0.2628 \\
  & Syn + Real & 25.63 & 0.9260 & 0.0833 & 164.37 & 0.0226 & 214.84 & 0.0411 & 0.2629 \\
\midrule
\multirow{2}{*}{\makecell{Light\\Tuning}}
  & LoRA (rank 32 / 128)  & 21.88 / 22.12 & 0.9012 / 0.9033 & 0.1027 / 0.1014 & 151.97 / 155.65 & 0.0157 / 0.0168 & 204.80 / 207.69 & 0.0324 / 0.0339 & \textbf{0.2682 / 0.2680} \\
  & Linear Adapters (rank 32 / 128) & 22.49 / 22.87 & 0.9069 / 0.9088 & 0.0959 / 0.0967 & 154.16 / 156.07 & 0.0186 / 0.0189 & 210.44 / 213.50 & 0.0372 / 0.0379 & 0.2678 / 0.2671 \\
\midrule
\multirow{1}{*}{\makecell{Train Free}} 
  & SDEdit & 22.21 & 0.9013 & 0.1169 & 147.06 & 0.0149 & 204.35 & 0.0327 & 0.2672 \\
\midrule
\multirow{5}{*}{Adapters} 
  & Domain Adapter (rank 32 / 128) & 25.61 / 25.60 & 0.9266 / 0.9264 & 0.0843 / 0.0848 & 164.17 / 163.64 & 0.0223 / 0.0220 & 215.80 / 215.28 & 0.0408 / 0.0398 & 0.2610 / 0.2608 \\
  & Spatial Adapter (rank 32 / 128) & 24.91 / 24.86 & 0.9200 / 0.9198 & 0.0858 / 0.0863 & 155.71 / 153.35 & 0.0197 / 0.0181 & 210.62 / 208.64 & 0.0393 / 0.0343 & 0.2634 / 0.2636 \\
  & Domain Switcher (2-Stage) & 24.94 & 0.9193 & 0.0888 & 157.89 & 0.0185 & 210.18 & 0.0350 & 0.2644 \\
  & Domain Switcher (Joint) & 23.16 & 0.9069 & 0.0991 & 145.48 & 0.0156 & 205.08 & 0.0335 & 0.2663 \\
  & Ours & 24.78 & 0.9153 & 0.0865 & \textbf{141.90} & \textbf{0.0121} & \textbf{200.24} & \textbf{0.0291} & 0.2674 \\
\bottomrule
\end{tabular}%
}
\end{table*}

%%%%%%%%% BOOKSTAB
\begin{table*}
\centering
\caption{\textbf{Ablation Study.} (top rows) The impact of our two-stage training and using real data at Stage 2.  (bottom rows) The contribution of \method components: Layer-Aware Training (LA Train), Domain Reassignment (Reassign), and inference shifting (Sampling).}
\vspace{-0.5em}
\label{eval:texturing_ablation}
\resizebox{\textwidth}{!}{%
\begin{tabular}{c c c c c c c c c c c}
\toprule
\multirow{2}{*}{\#} &
\multirow{2}{*}{Method} &
\multirow{2}{*}{\makecell{Training}} &
\multicolumn{3}{c}{\cellcolor{cyan!20} 3D Consistency} & 
\multicolumn{2}{c}{\cellcolor{magenta!20} Prior Preservation} & 
\multicolumn{2}{c}{\cellcolor{orange!20} Real-World Realism} &
\multicolumn{1}{c}{\cellcolor{yellow!20} Text Align} \\
\cmidrule(lr){4-6} \cmidrule(lr){7-8} \cmidrule(lr){9-10} \cmidrule(lr){11-11}
 & & & PSNR $\uparrow$ & SSIM $\uparrow$ & LPIPS $\downarrow$ & FID$_B$ $\downarrow$ & KID$_B$ $\downarrow$ &
 FID$_I$ $\downarrow$ & KID$_I$ $\downarrow$ & CLIP $\uparrow$ \\
\midrule
(1) & \makecell{DiT + \adaptershort} & \emph{Joint} & 25.53 & 0.9267 & 0.0850 & 166.93 & 0.0234 & 216.63 & 0.0424 & 0.2624 \\
(2) & \makecell{DiT + \adaptershort (w/o real data at Stage 2)} & 2-Stage & 25.11 & 0.9232 & 0.0849 & 162.02 & 0.0220 & 212.86 & 0.0395 & 0.2656 \\
\arrayrulecolor{blue}\midrule
\arrayrulecolor{black}   
(3) & \makecell{DiT + \adaptershort} & 2-Stage & 23.97 & 0.9123 & 0.0897 & 137.23 & 0.0098 & 198.44 & 0.0258 & 0.2695 \\
(4) & \makecell{DiT + \adaptershort + Sampling} & 2-Stage & 24.25 & 0.9128 & 0.0883 & 141.99 & 0.0123 & 202.07 & 0.0293 & 0.2679 \\
(5) & \makecell{DiT + \adaptershort + Reassign} & 2-Stage & 24.18 & 0.9125 & 0.0877 & 139.21 & 0.0102 & 198.71 & 0.0265 & 0.2678 \\
(6) & \makecell{DiT + \adaptershort + Reassig + Sampling} & 2-Stage & 24.37 & 0.9130 & 0.0871 & 141.08 & 0.0117 & 199.33 & 0.0289 & 0.2670 \\
(7) & \makecell{DiT + \adaptershort + LA Train + Reassign} & 2-Stage & 24.52 & 0.9141 & 0.0870 & 141.20 & 0.0118 & 199.39 & 0.0290 & 0.2675 \\
(8) & Ours (\adaptershort + LA Train + Reassign + Sampling) & 2-Stage & 24.78 & 0.9153 & 0.0865 & 141.90 & 0.0121 & 200.24 & 0.0291 & 0.2674 \\
\bottomrule
\end{tabular}%
}
\end{table*}

%\vspace{-0.5em}
\subsection{Text-to-Multiview Generation}

\paragraph{Task Definition.}

Given a set of $V\geq1$ camera viewpoints $\{c_{view}\}_0^V$, our goal is to generate 3D-consistent images $\{x_{RGB}\}_0^V$ matching the given views, thus learning the conditional distribution $q(\{x_{RGB}\}_0^V|\{c_{view}\}_0^V)$.

\noindent\textbf{Implementation Details}:
We generate images in a grid of four fixed viewpoints, where camera viewpoint information is injected via positional encoding into the latent samples.

\begin{table*}
\centering
\caption{\textbf{Text-to-Multiview Generation: Quantitative Results.} \method achieves significantly improved realism while maintaining 3D consistency comparable to the synthetic-only baseline (top row).}
\vspace{-0.5em}
\label{eval:t2m_quantitative}
\resizebox{0.9\textwidth}{!}{%
\begin{tabular}{c c c c c c c c c c}
\toprule
\multirow{2}{*}{} &
\multirow{2}{*}{Method} &
\multicolumn{3}{c}{\cellcolor{cyan!20} 3D Consistency} & 
\multicolumn{2}{c}{\cellcolor{magenta!20} Prior Preservation} & 
\multicolumn{2}{c}{\cellcolor{orange!20} Real-World Realism} &
\multicolumn{1}{c}{\cellcolor{yellow!20} Text Align} \\
\cmidrule(lr){3-5} \cmidrule(lr){6-7} \cmidrule(lr){8-9} \cmidrule(lr){10-10}
 & & PSNR $\uparrow$ & SSIM $\uparrow$ & LPIPS $\downarrow$ & FID$_B$ $\downarrow$ & KID$_B$ $\downarrow$ &
 FID$_I$ $\downarrow$ & KID$_I$ $\downarrow$ & CLIP $\uparrow$ \\
\midrule
\multirow{2}{*}{\makecell{Full\\Tuning}} 
  & Syn Only & \textbf{19.66} & \textbf{0.8779} & \textbf{0.0964} & 168.60 & 0.0204 & 215.57 & 0.0363 & 0.2541 \\
  & Syn + Real & 19.37 & 0.8714 & 0.1017 & 164.44 & 0.0192 & 214.72 & 0.0361 & 0.2586 \\
\midrule
\multirow{2}{*}{\makecell{Light\\Tuning}}
  & LoRA (rank 32 / 128)  & 17.54 / 17.91 & 0.8548 / 0.8568 & 0.1222 / 0.1186 & 153.89 / 154.11 & 0.0125 / 0.0131 & 204.14 / 206.68 & 0.0264 / 0.0269 & 0.2563 / 0.2579 \\
  & Linear Adapters (rank 32 / 128) & 18.07 / 18.38 & 0.8603 / 0.8690 & 0.1223 / 0.1191 & 152.21 / 155.40 & 0.0127 / 0.0150 & 204.02 / 208.78 & 0.0256 / 0.03133 & 0.2583 / 0.2526 \\
\midrule
\multirow{1}{*}{\makecell{Train Free}} 
  & SDEdit & 18.46 & 0.8431 & 0.1232 & 139.18 & 0.0112 & 199.67 & 0.0263 & 0.2609 \\
\midrule
\multirow{1}{*}{\makecell{Pretrained}} 
  & TRELLIS (large) & - & - & - & 181.92 & 0.0275 & 224.22 & 0.0441 & 0.2495 \\
\midrule
\multirow{5}{*}{Adapters} 
  & Domain Adapter (rank 32 / 128) & 19.27 / 19.02 & 0.8655 / 0.8621 & 0.1076 / 0.1082 & 158.52 / 157.45 & 0.0164 / 0.0163 & 212.00 / 210.67 & 0.0308 / 0.0313 & 0.2607 / 0.2601 \\
  & Spatial Adapter (Rank 32 / 128) & 18.48 / 18.36 & 0.8422 / 0.8350 & 0.1216 / 0.1234 & 132.48 / 130.25 & 0.0072 / 0.0071 & 200.38 / 199.46 & 0.0212 / 0.0205 & 0.2633 / 0.2638 \\
  & Domain Switcher (2-Stage) & 18.56 & 0.8481 & 0.1134 & 153.36 & 0.0146 & 208.14 & 0.0303 & 0.2587 \\
  & Domain Switcher (Joint) & 17.67 & 0.8010 & 0.1290 & 129.56 & 0.0061 & 197.11 & 0.0180 & 0.2629 \\
  & Ours & 19.02 & 0.8631 & 0.1075 & \textbf{122.01} & \textbf{0.0056} & \textbf{196.01} & \textbf{0.0171} & \textbf{0.2643} \\
\bottomrule
\end{tabular}%
}
\end{table*}

\begin{figure*}[h!]
\centering
\centerline{
    \includegraphics[width=0.9\linewidth]{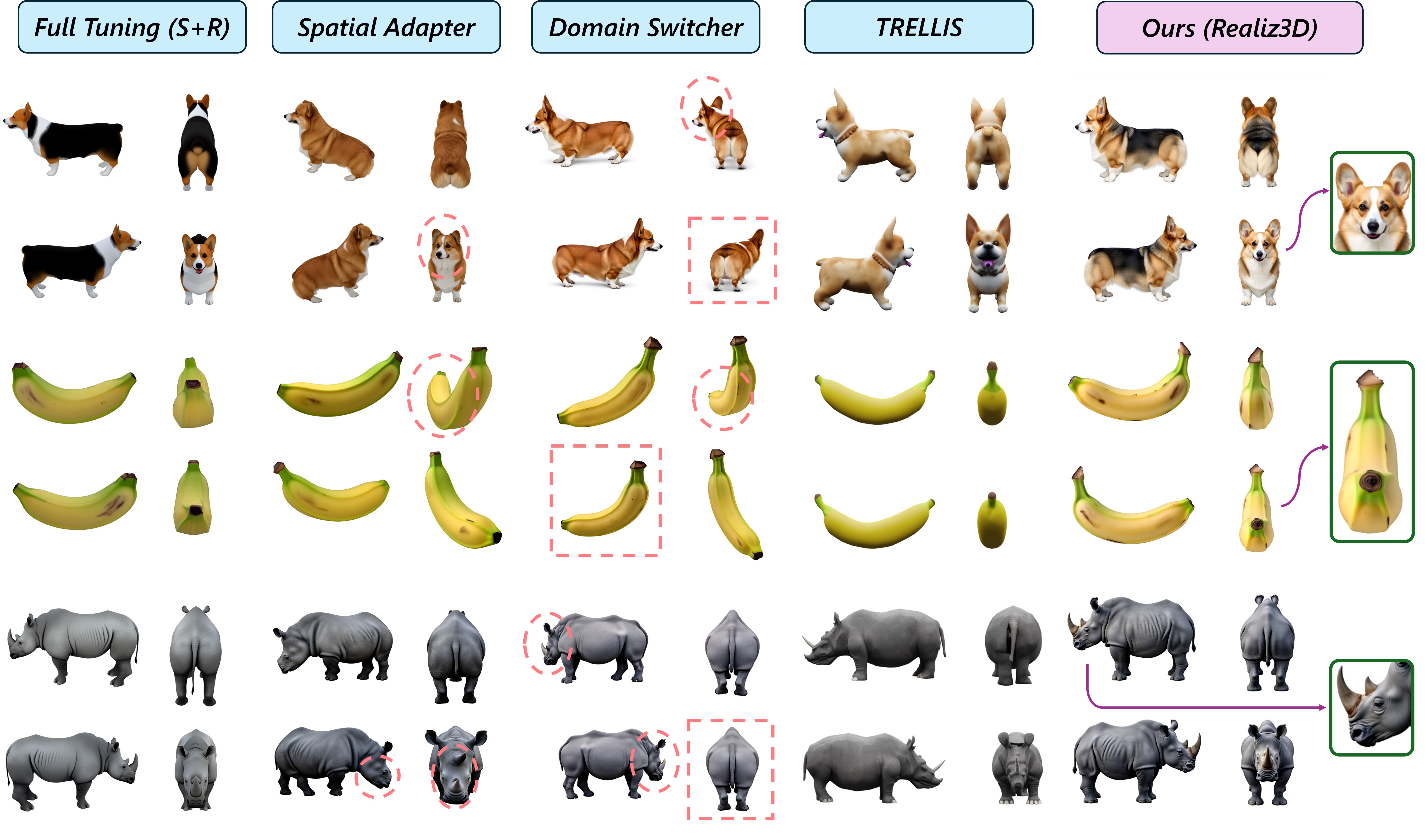}
}
\vspace{-0.5em}
\caption{\textbf{Text-to-Multiview: Qualitative Results.} ``A cute corgi dog'', ``A delicious ripe banana'', ``A rhino with thick gray skin'', each prompt was appended with ``highly photorealistic and detailed''. Red circles/squares highlight inconsistent regions/incorrect viewpoints, respectively. \method achieves notable improvements in photorealism while maintaining strong 3D consistency. Best viewed zoomed in.}
\label{fig:t2mv_baselines}
\end{figure*}

\noindent \textbf{Metrics.}
\noindent \textit{3D Consistency:}
We lift our generated multiview to 3D using a pre-trained LGM model~\cite{tang2024lgm}.
We measure reprojection error and report PSNR, SSIM and LPIPS.
\textit{Prior Preservation:} $FID_B$ and $KID_B$.
\textit{Real-World Realism:} $FID_I$ and $KID_I$.
\textit{Text-Image Alignment:}
CLIP score.

\noindent \textbf{Quantitative Evaluation.}
Main results appear in \cref{eval:t2m_quantitative}, using the prompts from the \textit{Evaluation Data} described earlier. While \method significantly improves realism, its 3D consistency remains slightly lower than fully synthetic baselines. As with texturing, 3D consistency is sensitive to fine-grained details, and lighting bias in the base model may cause inconsistencies. For pretrained models, we report realism-related metrics in~\cref{eval:t2m_quantitative}. Their performance is comparable or worse than our fully synthetic baseline, validating our conclusions, and the effectiveness of \method.

\noindent \textbf{Qualitative Evaluation.}
We present visual examples in \cref{fig:teaser} and \cref{fig:t2mv_baselines}. Only best-performing and external baselines are shown in the main paper. Full comparisons and additional results appear in the Appendix.
\section{Conclusions, Limitations and Future Work}
\vspace{-0.5em}
We introduced \textit{\method}, a fine-tuning strategy for diffusion models that enables controllability from synthetic data while preserving photorealism of the base model. \method is built on three key innovations: \textit{\adapterfull{s}}, designed to learn separable visual domains; \textit{layer-aware training} process, which maintains realism without sacrificing controllability; and \textit{domain reassignment}, which improves control transfer to real domain. A domain-aware sampling process boosts performance at test time.
Our work marks a significant step toward photorealistic 3D generation.
\textbf{Limitations.} 
While \method significantly improves realism, a small gap in control adherence remains (\cref{sec:experiments}). Key factors are: (1) 3D consistency is sensitive to fine-grained details; (2) Domain gaps can occasionally manifest in geometry, not just appearance; (3) The base model’s lighting bias can cause inconsistent appearance (failure cases are in the Appendix). Advances in relighting~\cite{liang2025diffusion, chaturvedi2025synthlight, litman2025lightswitch} open promising avenues to address this.
In addition, \method is currently designed to support control signals that are largely domain-agnostic, as text or geometric signals. This may limit direct applicability to other types of conditions, as images.
\clearpage
\section*{Acknowledgements}
We sincerely thank Timur Bagautdinov, Oran Gafni, Ita Lifshitz and Thu Nguyen-Phuoc for invaluable discussions and feedback. Or Litany acknowledges support from the Israel Science Foundation (grant 624/25) and the Azrieli Foundation Early Career Faculty Fellowship. This research was also supported in part by an academic gift from Meta. The authors gratefully acknowledge this support.

{
    \small
    \bibliographystyle{ieeenat_fullname}
    \bibliography{main}

\begin{thebibliography}{46}
\providecommand{\natexlab}[1]{#1}
\providecommand{\url}[1]{\texttt{#1}}
\expandafter\ifx\csname urlstyle\endcsname\relax
  \providecommand{\doi}[1]{doi: #1}\else
  \providecommand{\doi}{doi: \begingroup \urlstyle{rm}\Url}\fi

\bibitem[Amir et~al.(2021)Amir, Gandelsman, Bagon, and Dekel]{amir2021deep}
Shir Amir, Yossi Gandelsman, Shai Bagon, and Tali Dekel.
\newblock Deep vit features as dense visual descriptors.
\newblock \emph{arXiv preprint arXiv:2112.05814}, 2\penalty0 (3):\penalty0 4, 2021.

\bibitem[Bensadoun et~al.(2024)Bensadoun, Kleiman, Azuri, Harosh, Vedaldi, Neverova, and Gafni]{bensadoun2024meta}
Raphael Bensadoun, Yanir Kleiman, Idan Azuri, Omri Harosh, Andrea Vedaldi, Natalia Neverova, and Oran Gafni.
\newblock Meta 3d texturegen: Fast and consistent texture generation for 3d objects.
\newblock \emph{arXiv preprint arXiv:2407.02430}, 2024.

\bibitem[Chaturvedi et~al.(2025)Chaturvedi, Ren, Hold-Geoffroy, Liu, Dorsey, and Shu]{chaturvedi2025synthlight}
Sumit Chaturvedi, Mengwei Ren, Yannick Hold-Geoffroy, Jingyuan Liu, Julie Dorsey, and Zhixin Shu.
\newblock Synthlight: Portrait relighting with diffusion model by learning to re-render synthetic faces.
\newblock In \emph{Proceedings of the Computer Vision and Pattern Recognition Conference}, pages 369--379, 2025.

\bibitem[Chefer et~al.(2024)Chefer, Zada, Paiss, Ephrat, Tov, Rubinstein, Wolf, Dekel, Michaeli, and Mosseri]{chefer2024still}
Hila Chefer, Shiran Zada, Roni Paiss, Ariel Ephrat, Omer Tov, Michael Rubinstein, Lior Wolf, Tali Dekel, Tomer Michaeli, and Inbar Mosseri.
\newblock Still-moving: Customized video generation without customized video data.
\newblock \emph{ACM Transactions on Graphics (TOG)}, 43\penalty0 (6):\penalty0 1--11, 2024.

\bibitem[Daras et~al.(2023)Daras, Shah, Dagan, Gollakota, Dimakis, and Klivans]{daras2023ambient}
Giannis Daras, Kulin Shah, Yuval Dagan, Aravind Gollakota, Alex Dimakis, and Adam Klivans.
\newblock Ambient diffusion: Learning clean distributions from corrupted data.
\newblock In \emph{NeurIPS}, 2023.

\bibitem[Deitke et~al.(2023)Deitke, Schwenk, Salvador, Weihs, Michel, VanderBilt, Schmidt, Ehsani, Kembhavi, and Farhadi]{deitke2023objaverse}
Matt Deitke, Dustin Schwenk, Jordi Salvador, Luca Weihs, Oscar Michel, Eli VanderBilt, Ludwig Schmidt, Kiana Ehsani, Aniruddha Kembhavi, and Ali Farhadi.
\newblock Objaverse: A universe of annotated 3d objects.
\newblock In \emph{Proceedings of the IEEE/CVF Conference on Computer Vision and Pattern Recognition}, pages 13142--13153, 2023.

\bibitem[Deitke et~al.(2024)Deitke, Liu, Wallingford, Ngo, Michel, Kusupati, Fan, Laforte, Voleti, Gadre, et~al.]{deitke2024objaverse}
Matt Deitke, Ruoshi Liu, Matthew Wallingford, Huong Ngo, Oscar Michel, Aditya Kusupati, Alan Fan, Christian Laforte, Vikram Voleti, Samir~Yitzhak Gadre, et~al.
\newblock Objaverse-xl: A universe of 10m+ 3d objects.
\newblock \emph{Advances in Neural Information Processing Systems}, 36, 2024.

\bibitem[Deng et~al.(2009)Deng, Dong, Socher, Li, Li, and Fei-Fei]{deng2009imagenet}
Jia Deng, Wei Dong, Richard Socher, Li-Jia Li, Kai Li, and Li Fei-Fei.
\newblock Imagenet: A large-scale hierarchical image database.
\newblock In \emph{2009 IEEE conference on computer vision and pattern recognition}, pages 248--255. IEEE, 2009.

\bibitem[Esser et~al.(2024)Esser, Kulal, Blattmann, Entezari, M{\"u}ller, Saini, Levi, Lorenz, Sauer, Boesel, et~al.]{esser2024scaling}
Patrick Esser, Sumith Kulal, Andreas Blattmann, Rahim Entezari, Jonas M{\"u}ller, Harry Saini, Yam Levi, Dominik Lorenz, Axel Sauer, Frederic Boesel, et~al.
\newblock Scaling rectified flow transformers for high-resolution image synthesis.
\newblock In \emph{Forty-first international conference on machine learning}, 2024.

\bibitem[Gatis(2025)]{rembg}
Daniel Gatis.
\newblock {Rembg}: A tool to remove image backgrounds.
\newblock \url{https://github.com/danielgatis/rembg}, 2025.
\newblock Accessed: 2025-10-15.

\bibitem[Ghiasi et~al.(2022)Ghiasi, Kazemi, Borgnia, Reich, Shu, Goldblum, Wilson, and Goldstein]{ghiasi2022vision}
Amin Ghiasi, Hamid Kazemi, Eitan Borgnia, Steven Reich, Manli Shu, Micah Goldblum, Andrew~Gordon Wilson, and Tom Goldstein.
\newblock What do vision transformers learn? a visual exploration.
\newblock \emph{arXiv preprint arXiv:2212.06727}, 2022.

\bibitem[Guo et~al.(2023)Guo, Yang, Rao, Liang, Wang, Qiao, Agrawala, Lin, and Dai]{guo2023animatediff}
Yuwei Guo, Ceyuan Yang, Anyi Rao, Zhengyang Liang, Yaohui Wang, Yu Qiao, Maneesh Agrawala, Dahua Lin, and Bo Dai.
\newblock Animatediff: Animate your personalized text-to-image diffusion models without specific tuning.
\newblock \emph{arXiv preprint arXiv:2307.04725}, 2023.

\bibitem[Ho and Salimans(2022)]{ho2022classifier}
Jonathan Ho and Tim Salimans.
\newblock Classifier-free diffusion guidance.
\newblock \emph{arXiv preprint arXiv:2207.12598}, 2022.

\bibitem[Hu et~al.(2021)Hu, Shen, Wallis, Allen-Zhu, Li, Wang, Wang, and Chen]{hu2021lora}
Edward~J Hu, Yelong Shen, Phillip Wallis, Zeyuan Allen-Zhu, Yuanzhi Li, Shean Wang, Lu Wang, and Weizhu Chen.
\newblock Lora: Low-rank adaptation of large language models.
\newblock \emph{arXiv preprint arXiv:2106.09685}, 2021.

\bibitem[Hu et~al.(2022)Hu, Shen, Wallis, Allen-Zhu, Li, Wang, Wang, Chen, et~al.]{hu2022lora}
Edward~J Hu, Yelong Shen, Phillip Wallis, Zeyuan Allen-Zhu, Yuanzhi Li, Shean Wang, Lu Wang, Weizhu Chen, et~al.
\newblock Lora: Low-rank adaptation of large language models.
\newblock \emph{ICLR}, 1\penalty0 (2):\penalty0 3, 2022.

\bibitem[Jiang et~al.(2025)Jiang, Wang, Li, Zhang, Wang, Wei, Dai, Zhang, and Wang]{jiang2025no}
Dengyang Jiang, Mengmeng Wang, Liuzhuozheng Li, Lei Zhang, Haoyu Wang, Wei Wei, Guang Dai, Yanning Zhang, and Jingdong Wang.
\newblock No other representation component is needed: Diffusion transformers can provide representation guidance by themselves.
\newblock \emph{arXiv preprint arXiv:2505.02831}, 2025.

\bibitem[Li et~al.(2023)Li, Liu, Wu, Mu, Yang, Gao, Li, and Lee]{li2023gligen}
Yuheng Li, Haotian Liu, Qingyang Wu, Fangzhou Mu, Jianwei Yang, Jianfeng Gao, Chunyuan Li, and Yong~Jae Lee.
\newblock Gligen: Open-set grounded text-to-image generation.
\newblock In \emph{Proceedings of the IEEE/CVF conference on computer vision and pattern recognition}, pages 22511--22521, 2023.

\bibitem[Liang et~al.(2025)Liang, Gojcic, Ling, Munkberg, Hasselgren, Lin, Gao, Keller, Vijaykumar, Fidler, et~al.]{liang2025diffusion}
Ruofan Liang, Zan Gojcic, Huan Ling, Jacob Munkberg, Jon Hasselgren, Chih-Hao Lin, Jun Gao, Alexander Keller, Nandita Vijaykumar, Sanja Fidler, et~al.
\newblock Diffusion renderer: Neural inverse and forward rendering with video diffusion models.
\newblock In \emph{Proceedings of the Computer Vision and Pattern Recognition Conference}, pages 26069--26080, 2025.

\bibitem[Lin et~al.(2025)Lin, Yang, Chen, Xu, Yan, Wu, Xu, Xu, Zhang, and Chen]{lin2025kiss3dgen}
Jiantao Lin, Xin Yang, Meixi Chen, Yingjie Xu, Dongyu Yan, Leyi Wu, Xinli Xu, Lie Xu, Shunsi Zhang, and Ying-Cong Chen.
\newblock Kiss3dgen: Repurposing image diffusion models for 3d asset generation.
\newblock In \emph{Proceedings of the Computer Vision and Pattern Recognition Conference}, pages 5870--5880, 2025.

\bibitem[Litman et~al.(2025)Litman, De~la Torre, and Tulsiani]{litman2025lightswitch}
Yehonathan Litman, Fernando De~la Torre, and Shubham Tulsiani.
\newblock Lightswitch: Multi-view relighting with material-guided diffusion.
\newblock In \emph{Proceedings of the IEEE/CVF International Conference on Computer Vision}, pages 27750--27759, 2025.

\bibitem[Liu et~al.(2023)Liu, Wu, Van~Hoorick, Tokmakov, Zakharov, and Vondrick]{liu2023zero}
Ruoshi Liu, Rundi Wu, Basile Van~Hoorick, Pavel Tokmakov, Sergey Zakharov, and Carl Vondrick.
\newblock Zero-1-to-3: Zero-shot one image to 3d object.
\newblock In \emph{Proceedings of the IEEE/CVF international conference on computer vision}, pages 9298--9309, 2023.

\bibitem[Long et~al.(2024)Long, Guo, Lin, Liu, Dou, Liu, Ma, Zhang, Habermann, Theobalt, et~al.]{long2024wonder3d}
Xiaoxiao Long, Yuan-Chen Guo, Cheng Lin, Yuan Liu, Zhiyang Dou, Lingjie Liu, Yuexin Ma, Song-Hai Zhang, Marc Habermann, Christian Theobalt, et~al.
\newblock Wonder3d: Single image to 3d using cross-domain diffusion.
\newblock In \emph{Proceedings of the IEEE/CVF Conference on Computer Vision and Pattern Recognition}, pages 9970--9980, 2024.

\bibitem[Luo et~al.(2023)Luo, Dunlap, Park, Holynski, and Darrell]{luo2023diffusion}
Grace Luo, Lisa Dunlap, Dong~Huk Park, Aleksander Holynski, and Trevor Darrell.
\newblock Diffusion hyperfeatures: Searching through time and space for semantic correspondence.
\newblock \emph{Advances in Neural Information Processing Systems}, 36:\penalty0 47500--47510, 2023.

\bibitem[Meng et~al.(2022)Meng, He, Song, Song, Wu, Zhu, and Ermon]{meng2021sdedit}
Chenlin Meng, Yutong He, Yang Song, Jiaming Song, Jiajun Wu, Jun-Yan Zhu, and Stefano Ermon.
\newblock {SDEdit:} guided image synthesis and editing with stochastic differential equations.
\newblock In \emph{ICLR}, 2022.

\bibitem[Mou et~al.(2024)Mou, Wang, Xie, Wu, Zhang, Qi, and Shan]{mou2024t2i}
Chong Mou, Xintao Wang, Liangbin Xie, Yanze Wu, Jian Zhang, Zhongang Qi, and Ying Shan.
\newblock T2i-adapter: Learning adapters to dig out more controllable ability for text-to-image diffusion models.
\newblock In \emph{Proceedings of the AAAI conference on artificial intelligence}, pages 4296--4304, 2024.

\bibitem[Peebles and Xie(2023)]{peebles2023scalable}
William Peebles and Saining Xie.
\newblock Scalable diffusion models with transformers.
\newblock In \emph{Proceedings of the IEEE/CVF International Conference on Computer Vision}, pages 4195--4205, 2023.

\bibitem[Peng et~al.(2024)Peng, Sobol, Tomizuka, Keutzer, Xu, and Litany]{peng2024lesson}
Chensheng Peng, Ido Sobol, Masayoshi Tomizuka, Kurt Keutzer, Chenfeng Xu, and Or Litany.
\newblock A lesson in splats: Teacher-guided diffusion for 3d gaussian splats generation with 2d supervision.
\newblock \emph{arXiv preprint arXiv:2412.00623}, 2024.

\bibitem[Radford et~al.(2021)Radford, Kim, Hallacy, Ramesh, Goh, Agarwal, Sastry, Askell, Mishkin, Clark, et~al.]{radford2021learning}
Alec Radford, Jong~Wook Kim, Chris Hallacy, Aditya Ramesh, Gabriel Goh, Sandhini Agarwal, Girish Sastry, Amanda Askell, Pamela Mishkin, Jack Clark, et~al.
\newblock Learning transferable visual models from natural language supervision.
\newblock In \emph{International conference on machine learning}, pages 8748--8763. PMLR, 2021.

\bibitem[Rebuffi et~al.(2017)Rebuffi, Bilen, and Vedaldi]{rebuffi17learning}
Sylvestre{-}Alvise Rebuffi, Hakan Bilen, and Andrea Vedaldi.
\newblock Learning multiple visual domains with residual adapters.
\newblock In \emph{NeurIPS}, 2017.

\bibitem[Ronneberger et~al.(2015)Ronneberger, Fischer, and Brox]{ronneberger2015u}
Olaf Ronneberger, Philipp Fischer, and Thomas Brox.
\newblock U-net: Convolutional networks for biomedical image segmentation.
\newblock In \emph{Medical image computing and computer-assisted intervention--MICCAI 2015: 18th international conference, Munich, Germany, October 5-9, 2015, proceedings, part III 18}, pages 234--241. Springer, 2015.

\bibitem[Shi et~al.(2023)Shi, Chen, Zhang, Liu, Xu, Wei, Chen, Zeng, and Su]{shi2023zero123++}
Ruoxi Shi, Hansheng Chen, Zhuoyang Zhang, Minghua Liu, Chao Xu, Xinyue Wei, Linghao Chen, Chong Zeng, and Hao Su.
\newblock Zero123++: a single image to consistent multi-view diffusion base model.
\newblock \emph{arXiv preprint arXiv:2310.15110}, 2023.

\bibitem[Shi et~al.(2024)Shi, Wang, Ye, Long, Li, and Yang]{mvdream}
Yichun Shi, Peng Wang, Jianglong Ye, Mai Long, Kejie Li, and Xiao Yang.
\newblock Mvdream: Multi-view diffusion for 3d generation, 2024.

\bibitem[Sobol et~al.(2024)Sobol, Xu, and Litany]{sobol2024zero}
Ido Sobol, Chenfeng Xu, and Or Litany.
\newblock Zero-to-hero: Enhancing zero-shot novel view synthesis via attention map filtering.
\newblock \emph{Advances in Neural Information Processing Systems}, 37:\penalty0 30522--30553, 2024.

\bibitem[Song et~al.(2020)Song, Meng, and Ermon]{song2020denoising}
Jiaming Song, Chenlin Meng, and Stefano Ermon.
\newblock Denoising diffusion implicit models.
\newblock \emph{arXiv preprint arXiv:2010.02502}, 2020.

\bibitem[Song et~al.(2021)Song, Meng, and Ermon]{song2021ddim}
Jiaming Song, Chenlin Meng, and Stefano Ermon.
\newblock Denoising diffusion implicit models.
\newblock In \emph{International Conference on Learning Representations}, 2021.

\bibitem[Stern et~al.(2025)Stern, Sobol, and Litany]{stern2025appreciate}
Saar Stern, Ido Sobol, and Or Litany.
\newblock Appreciate the view: A task-aware evaluation framework for novel view synthesis.
\newblock \emph{arXiv preprint arXiv:2511.12675}, 2025.

\bibitem[Tang et~al.(2024)Tang, Chen, Chen, Wang, Zeng, and Liu]{tang2024lgm}
Jiaxiang Tang, Zhaoxi Chen, Xiaokang Chen, Tengfei Wang, Gang Zeng, and Ziwei Liu.
\newblock Lgm: Large multi-view gaussian model for high-resolution 3d content creation.
\newblock \emph{arXiv preprint arXiv:2402.05054}, 2024.

\bibitem[Tumanyan et~al.(2023)Tumanyan, Geyer, Bagon, and Dekel]{tumanyan2023plug}
Narek Tumanyan, Michal Geyer, Shai Bagon, and Tali Dekel.
\newblock Plug-and-play diffusion features for text-driven image-to-image translation.
\newblock In \emph{Proceedings of the IEEE/CVF Conference on Computer Vision and Pattern Recognition}, pages 1921--1930, 2023.

\bibitem[Wang et~al.(2004)Wang, Bovik, Sheikh, and Simoncelli]{ssim_2004}
Zhou Wang, A.C. Bovik, H.R. Sheikh, and E.P. Simoncelli.
\newblock Image quality assessment: from error visibility to structural similarity.
\newblock \emph{IEEE Transactions on Image Processing}, 13\penalty0 (4):\penalty0 600--612, 2004.

\bibitem[Xiang et~al.(2024)Xiang, Lv, Xu, Deng, Wang, Zhang, Chen, Tong, and Yang]{Xiang2024Structured3L}
Jianfeng Xiang, Zelong Lv, Sicheng Xu, Yu Deng, Ruicheng Wang, Bowen Zhang, Dong Chen, Xin Tong, and Jiaolong Yang.
\newblock Structured 3d latents for scalable and versatile 3d generation.
\newblock \emph{2025 IEEE/CVF Conference on Computer Vision and Pattern Recognition (CVPR)}, pages 21469--21480, 2024.

\bibitem[Xu et~al.(2025)Xu, Ge, Lin, Feng, Xu, Zhao, Zhang, and Chen]{xu2025flexgen}
Xinli Xu, Wenhang Ge, Jiantao Lin, Jiawei Feng, Lie Xu, HanFeng Zhao, Shunsi Zhang, and Ying-Cong Chen.
\newblock Flexgen: Flexible multi-view generation from text and image inputs.
\newblock In \emph{Proceedings of the IEEE/CVF International Conference on Computer Vision}, pages 18714--18724, 2025.

\bibitem[Yang et~al.(2023)Yang, Zhou, Feng, and Wang]{yang2023diffusion}
Xingyi Yang, Daquan Zhou, Jiashi Feng, and Xinchao Wang.
\newblock Diffusion probabilistic model made slim.
\newblock In \emph{Proceedings of the IEEE/CVF Conference on computer vision and pattern recognition}, pages 22552--22562, 2023.

\bibitem[Yi et~al.(2024)Yi, Li, Xin, and Li]{yi2024towards}
Mingyang Yi, Aoxue Li, Yi Xin, and Zhenguo Li.
\newblock Towards understanding the working mechanism of text-to-image diffusion model.
\newblock \emph{arXiv preprint arXiv:2405.15330}, 2024.

\bibitem[Zhang et~al.(2024)Zhang, Bi, Tan, Xiangli, Zhao, Sunkavalli, and Xu]{gslrm2024}
Kai Zhang, Sai Bi, Hao Tan, Yuanbo Xiangli, Nanxuan Zhao, Kalyan Sunkavalli, and Zexiang Xu.
\newblock Gs-lrm: Large reconstruction model for 3d gaussian splatting.
\newblock \emph{European Conference on Computer Vision}, 2024.

\bibitem[Zhang et~al.(2023)Zhang, Rao, and Agrawala]{zhang2023adding}
Lvmin Zhang, Anyi Rao, and Maneesh Agrawala.
\newblock Adding conditional control to text-to-image diffusion models.
\newblock In \emph{Proceedings of the IEEE/CVF International Conference on Computer Vision}, pages 3836--3847, 2023.

\bibitem[Zhang et~al.(2018)Zhang, Isola, Efros, Shechtman, and Wang]{lpips_cvpr_2018}
Richard Zhang, Phillip Isola, Alexei~A. Efros, Eli Shechtman, and Oliver Wang.
\newblock The unreasonable effectiveness of deep features as a perceptual metric, 2018.

\end{thebibliography}
}
\clearpage
\appendix
\section*{Appendix}
\label{sec:appendix}

\startcontents[appendices]
\printcontents[appendices]{}{1}{\subsection*{Contents}}

\section{Feature Maps in Diffusion Models}
\subsection{Motivation}
\label{subset:feature_motivation}
As thoroughly discussed in the \textit{Diffusion Models and Domain Gaps} section (Section 3) of the main paper, prior works has shown that visual details evolve throughout the denoising process in diffusion models, both across timesteps and across layers of the neural network. Building on studies that analyze diffusion features~\cite{tumanyan2023plug, jiang2025no}, we examine and visualize the internal representations of our diffusion transformer backbone over the course of denoising.

To do so, we first generate diverse prompts describing random objects in random styles (e.g., photorealistic, cartoon, watercolor, anime, comic, etc.) using an LLM. We then synthesize images from these prompts using our base DiT text-to-image model, running generation with 20 DDIM steps.

\begin{figure*}
\centering
\begin{minipage}{0.55\textwidth}
    \centering
    \includegraphics[width=\linewidth]{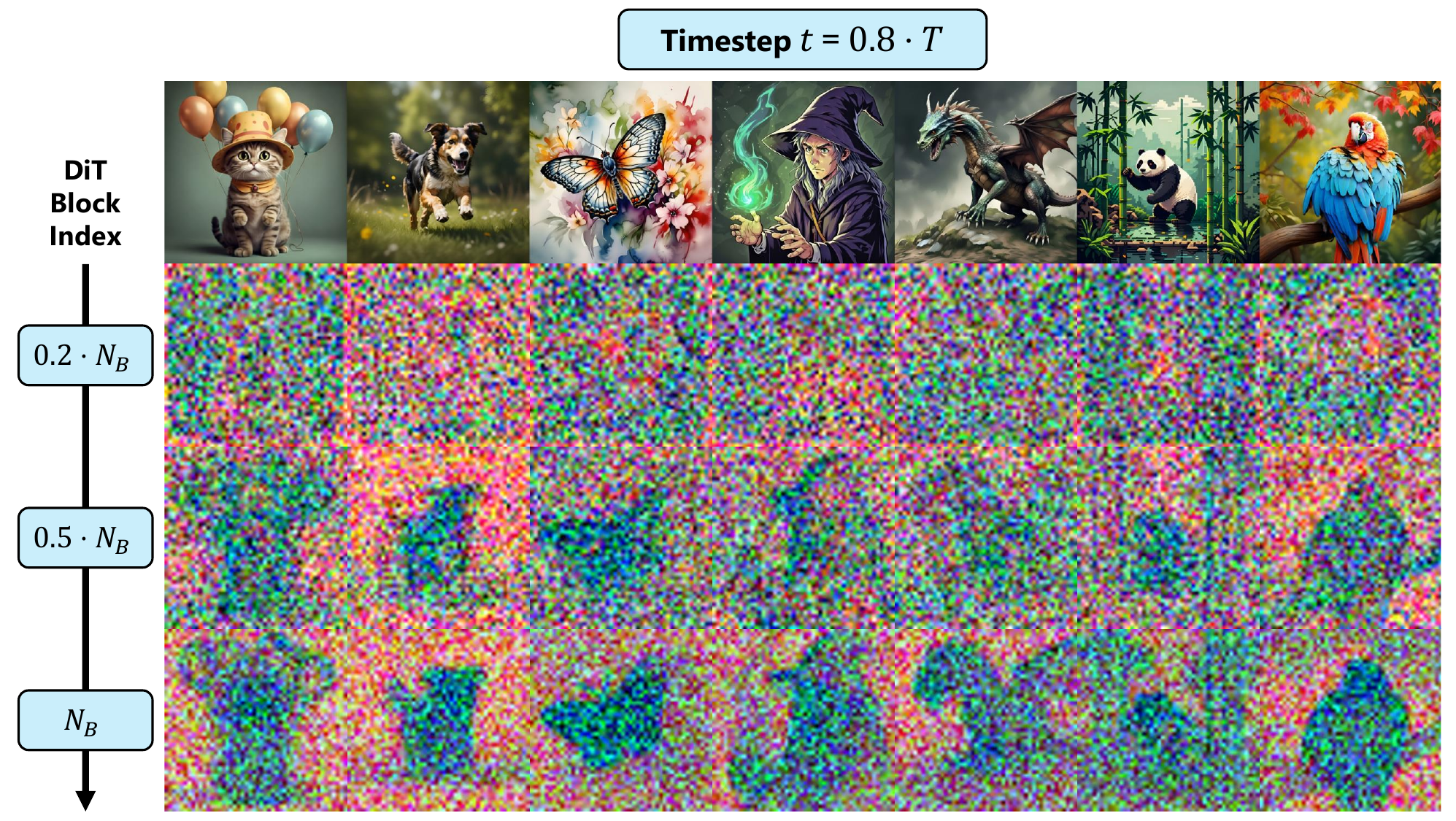}
\end{minipage}

\begin{minipage}{0.55\textwidth}
    \centering
    \includegraphics[width=\linewidth]{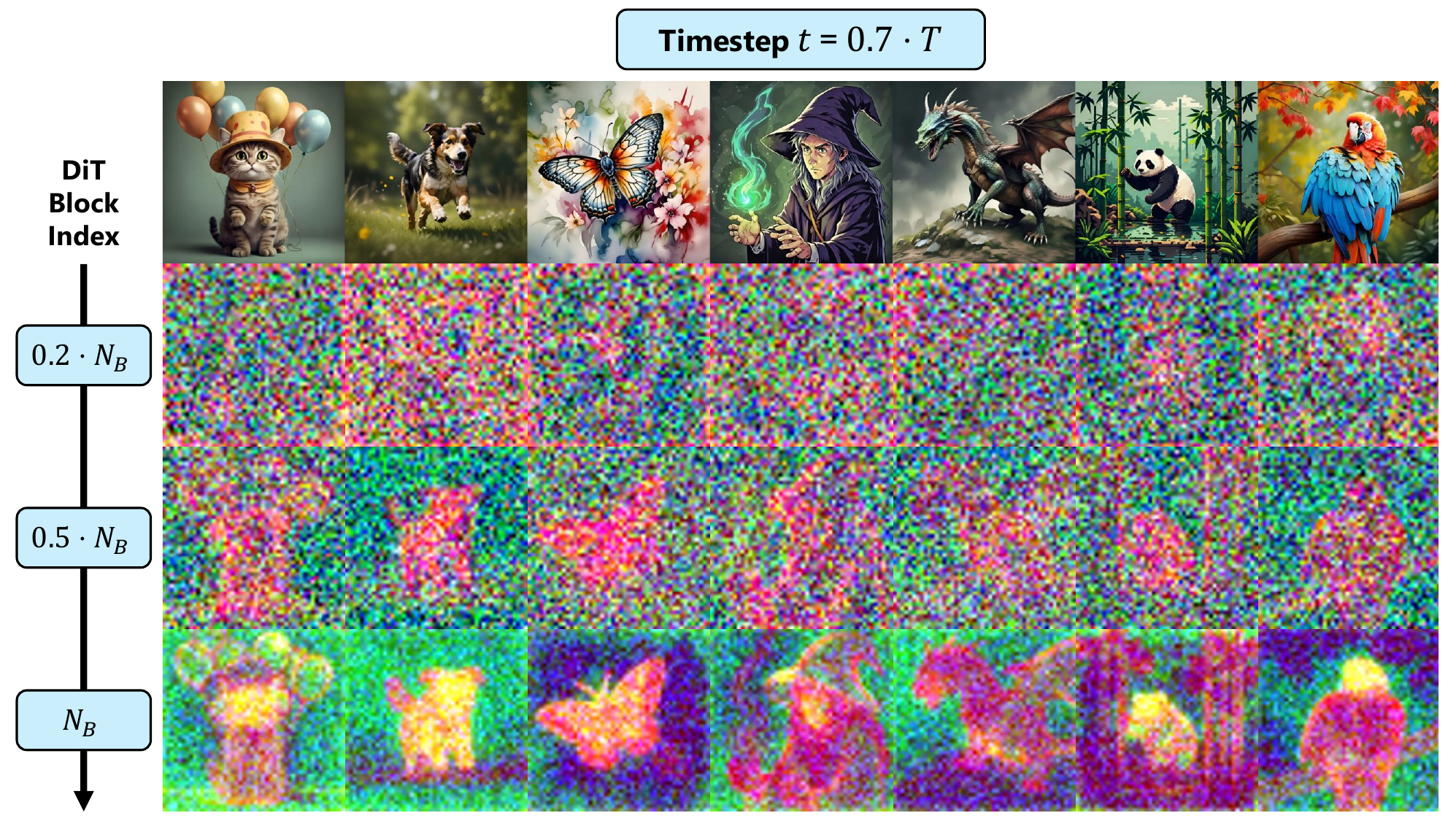}
\end{minipage}

\begin{minipage}{0.55\textwidth}
    \centering
    \includegraphics[width=\linewidth]{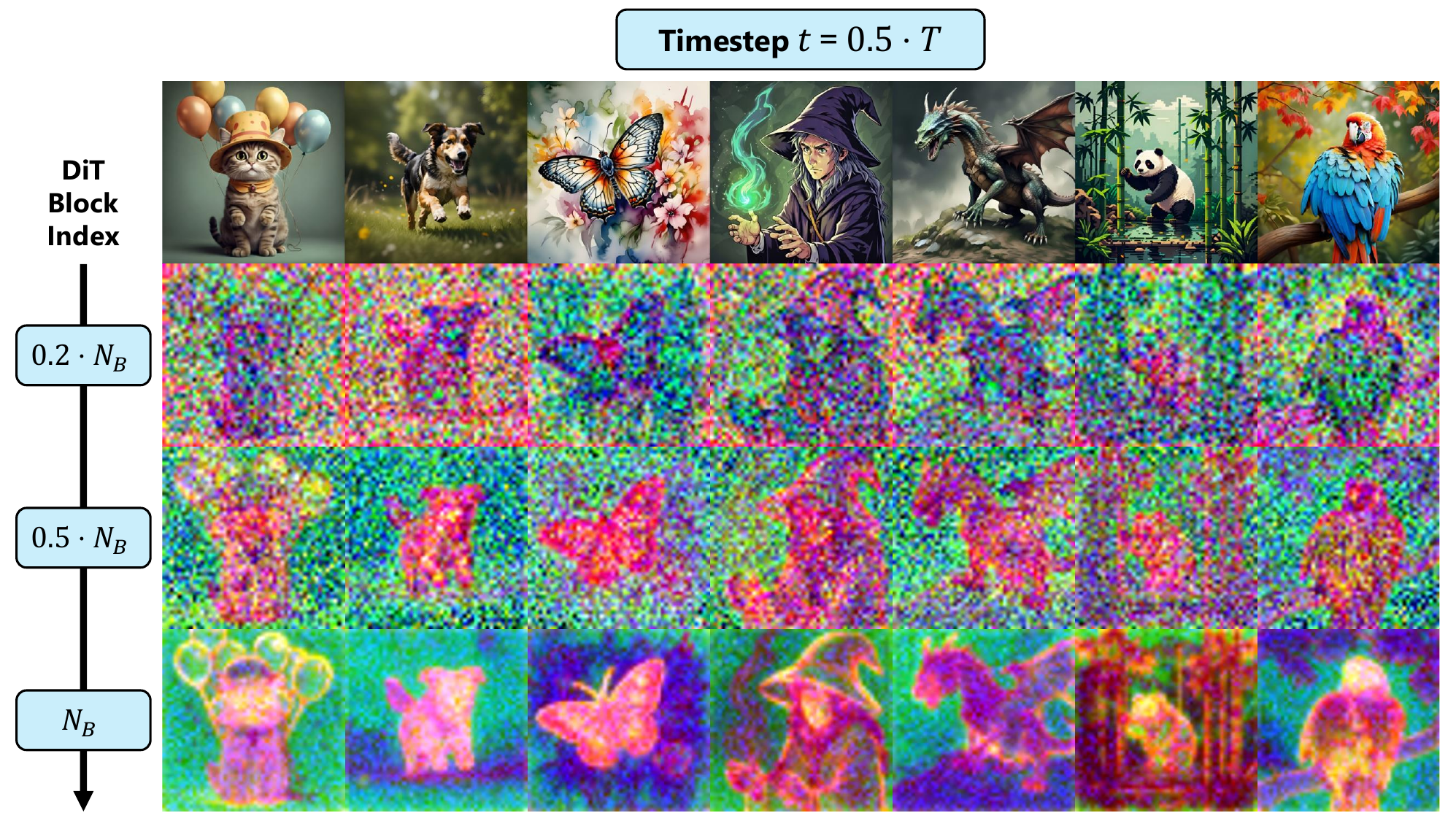}
\end{minipage}

\begin{minipage}{0.55\textwidth}
    \centering
    \includegraphics[width=\linewidth]{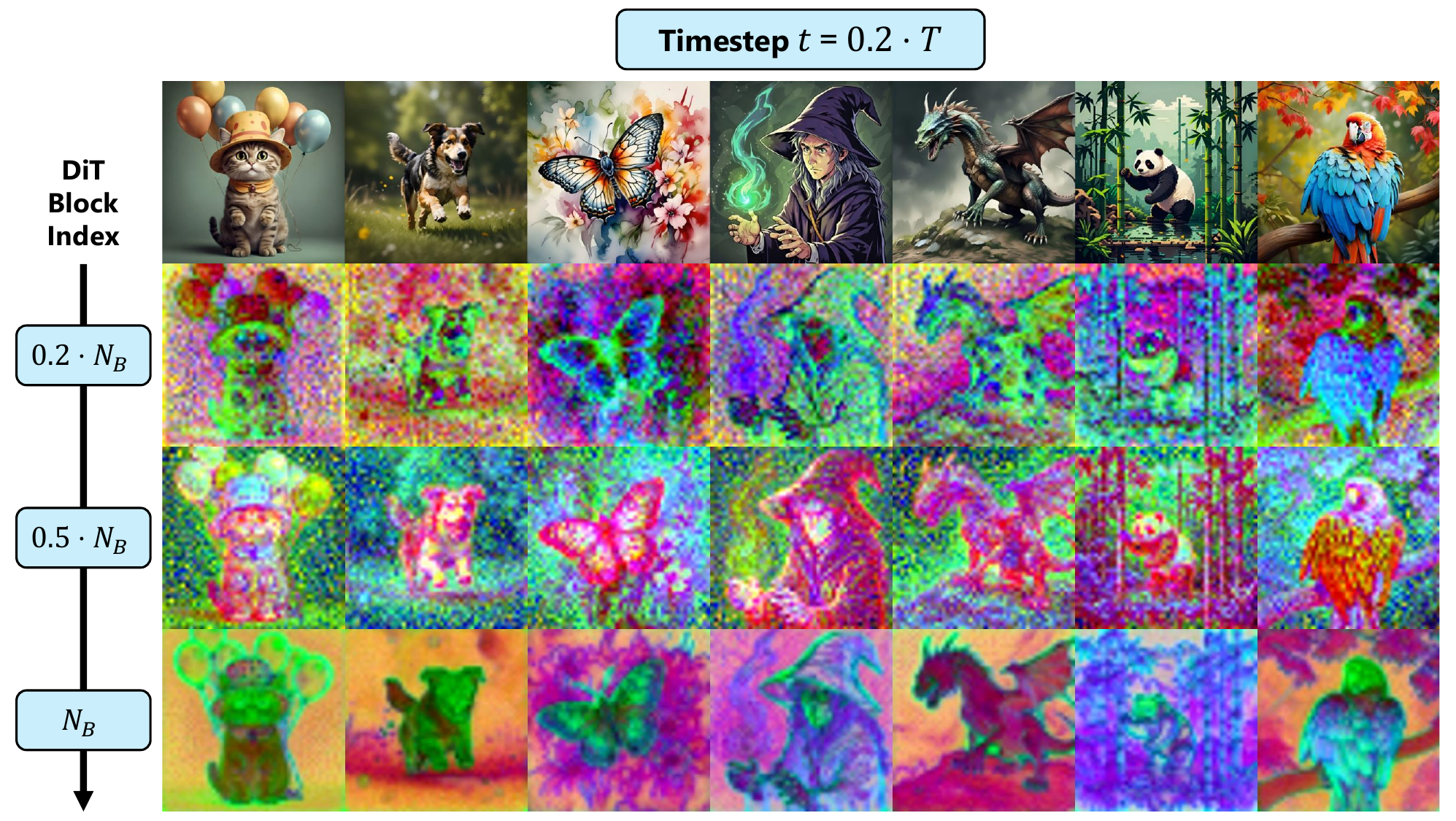}
\end{minipage}
\caption{Diffusion features, extracted from our base T2I model at different timesteps and layers during the generation process. We visualize the first three PCA components as RGB images. See Sec.~\ref{subset:feature_motivation} for additional details and analysis.}
\label{fig:dit_features}
\end{figure*}

During generation, we extract intermediate feature maps at multiple timesteps and layers of the network. We apply PCA to these features and visualize the resulting components as RGB images, presented in~\figref{fig:dit_features}. Importantly, PCA is computed independently for each timestep–layer pair, so the colors are not consistent across timesteps or layers.
Concretely, we extract features at four timesteps (800, 700, 500, and 200, with $T = 1000$) and at three layers corresponding to the beginning, middle, and end of the model. Let $N_B$ denote the total number of DiT blocks.

We present the PCA-reduced features in~\figref{fig:dit_features}. The figure comprises four vertically stacked subfigures, where each subfigure shows features from different layers at a \textbf{fixed} timestep. Within each subfigure, rows follow the order of the network depth. The top row shows features from an early layer in the architecture, and the bottom row shows features from a later layer.
Similarly, the ordering of the subfigures follows the denoising schedule: the top subfigure corresponds to an early timestep at sampling (where the noise level is highest), and the bottom subfigure corresponds to a late timestep (where the noise level is lowest).

As shown in the figure, when the noise level is high, the features primarily capture coarse and structural information, which is largely domain-agnostic. As the noise level decreases, the features increasingly reflect high-frequency patterns and fine-grained details.

The denoising layers exhibit a similar trend: early layers encode coarse structural patterns, shared between synthetic and real data, whereas later layers capture fine details.

\subsection{Case Study: Layer-Selective Training in 2D Image Generation}
\label{subset:case_study}
As discussed above and on the main paper, early layers in the diffusion transformer network usually correspond to structural and coarse patterns, while later laters capture fine-details and high-frequency patterns. We leverage this observation in our proposed Layer-Aware Training, where real data affect later layer more strongly and synthetic data has a stronger influence of the early layers.

\begin{figure*}
    \centering
    \includegraphics[width=0.75\linewidth]{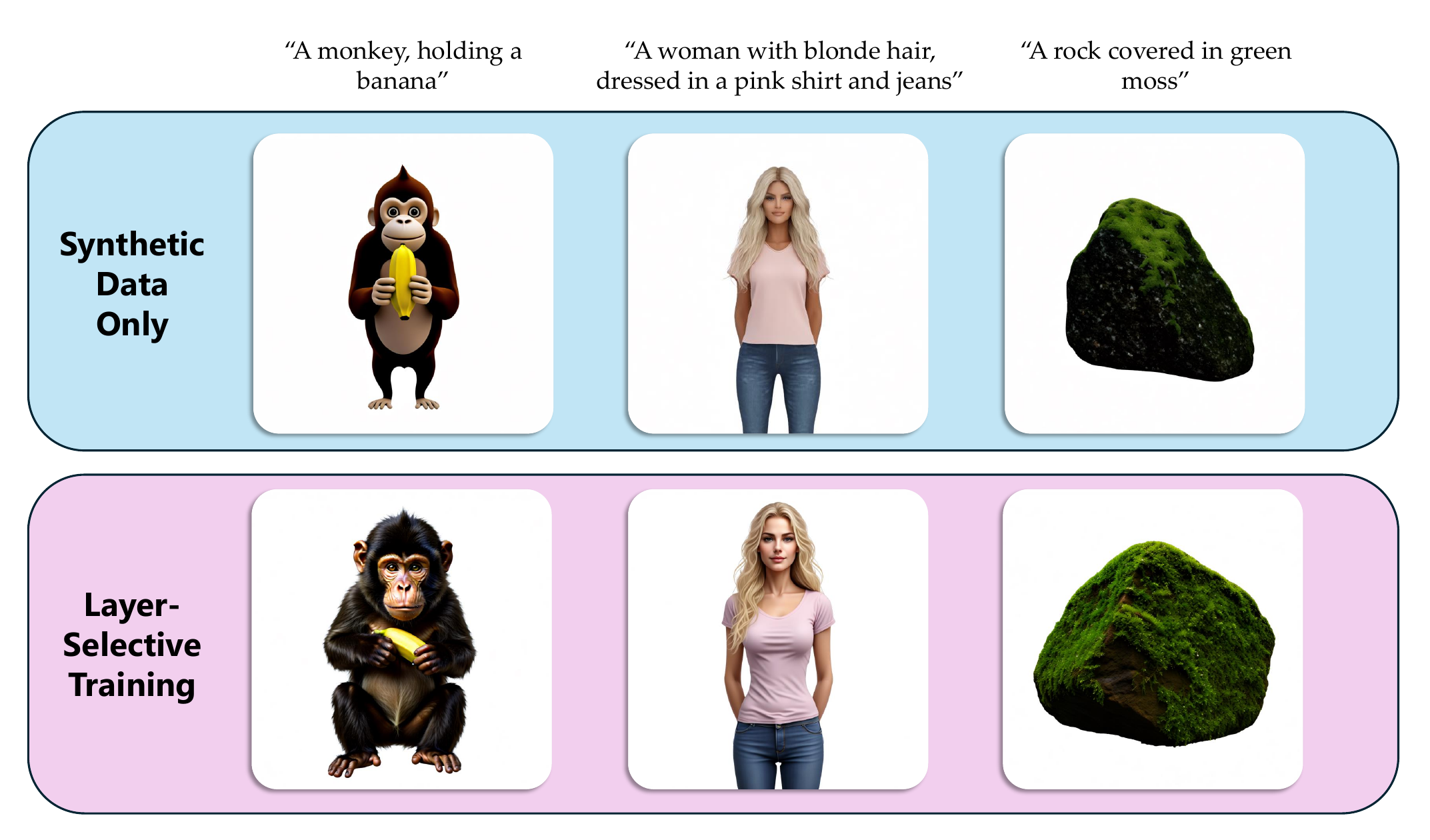}
\caption{\textbf{Layer-Selective Training.} Training early blocks on synthetic data and later blocks on real data enables the model to learn controllable properties from synthetic data while maintaining photorealism. See further details in Sec.~\ref{subset:case_study}.}
    \label{fig:layer_aware}
\end{figure*}

Before developing our Layer-Aware Training strategy, we conducted a proof-of-concept experiment for image generation, to test whether this insight could be used during training to gain controllability from synthetic data while preserving realism from real data. The results are shown in Fig.~\ref{fig:layer_aware}.

We conducted the proof-of-concept experiment using two equal-sized datasets: one synthetic and one realistic. The synthetic images were rendered from 3D assets, while the realistic images were generated by a T2I model using prompts describing similar objects, augmented with the phrases “highly realistic” and “white background” appended to each prompt.

We fine-tuned the base DiT T2I model in two ways. First, we fine-tuned it solely on synthetic data. Second, we applied a layer-selective strategy: the first 50\% of DiT blocks were fine-tuned on synthetic data (while the later blocks were frozen), and the remaining 50\% were fine-tuned on realistic data (while the early blocks were frozen). Although both approaches caused the model to shift toward object-centric generation with white backgrounds, the layer-Selective strategy consistently preserves the base model’s realistic prior more effectively than the full-synthetic baseline, as shown in Fig.~\ref{fig:layer_aware}. While a similar experiment was done with timestep-selective training, we found the layer-selective approach to be more stable and robust.

This proof-of-concept experiment provides the core motivation and inspiration for the layer-aware training approach described in the main paper.

\section{Method}
In this section, we provide additional information and implementation details about the different components of \method.
\subsection{\adapterfull{s} (Stage 1)}
\label{app:stage_1}
\paragraph{Discussion: \adapterfull Design}
Our design builds on the intuition that our goal is to rebalance existing visual modes within the pretrained model rather than introduce new modalities. Prior adapter-based methods~\cite{guo2023animatediff, long2024wonder3d} handle much larger modality shifts: AnimateDiff~\cite{guo2023animatediff} suppresses the low-resolution and noisy video latent pathway to inject temporal conditioning, while Wonder3D adds conditioning streams for normal maps that differ substantially from natural images. In our case, both real and synthetic imagery already occupy the model’s learned feature space, making a lightweight low-rank residual sufficient to adjust their balance without disrupting pretrained representations~\cite{rebuffi17learning}.

\subsection{Fine-tuning with Representation Binding (Stage 2)}
\label{app:stage_2}
As described in the main paper, we incorporate real samples during training to prevent forgetting realism. Since real images lack explicit control supervision, naïvely training on them could disrupt control learned from synthetic data. To avoid this, we use real data to mostly update the later diffusion blocks, responsible for appearance refinement, while freezing the early blocks.

Concretely, During each real-data training iteration, we freeze DiT blocks $B \in [0, B_i]$, where $i$ is an integer block index randomly drawn from $[0, \tau_B]$.
This stochastic layer-freezing regularizes more strongly early representations, without requiring a fixed cutoff.
We find the stochastic approach to be robust to small variations and selection of configuration. Empirically, setting $\tau_B \in [0.4, 0.5]$ of the total number of blocks provides stable and robust performance.
%\TODO{remove the numbers - add to appendix}

Throughout the entire training process, we rely solely on the traditional diffusion loss, which is well established and empirically stable. At stage 1, the diffusion objective is used to train our \adapterfull{s}, and at stage 2 the same objective is used to train the DiT backbone.
\noindent When training on synthetic samples, the training objective is 
$
\mathcal{L} 
= \mathbb{E}_{z,\epsilon \sim \mathcal{N}(0,1),\, t}
\left[
\left\lVert \epsilon - \epsilon_{\theta}(z_t, t, c, e_{syn}) \right\rVert_2^2
\right]
$ (at stage 1, $c=\varnothing$).

\noindent When training on realistic samples, the training objective is 
$
\mathcal{L} 
= \mathbb{E}_{z,\epsilon \sim \mathcal{N}(0,1),\, t}
\left[
\left\lVert \epsilon - \epsilon_{\theta}(z_t, t, c=\varnothing, e_{real}) \right\rVert_2^2
\right]
$.

\subsection{Inference-time Domain Shifting}
\label{app:inference}
As discussed in the main paper, to further promote control transfer, we introduce \textit{Domain Reassignment}.
With probability $p_B$, we reassign early DiT blocks ($B\in[0,B_j]$, $j$ is an integer, sampled from $[0, \tau_B]$) to operate in synthetic mode even when processing real samples; that is, we substitute $e_\text{domain}\leftarrow e_\text{syn}$ in the corresponding \adapterfull{}s.

\noindent In practice, we set $e_\text{domain}=e_\text{syn}$ when either of the following conditions hold:
\begin{itemize}[leftmargin=*]
\item \textit{Layer-based shifting:} For all timesteps $t$ and DiT blocks $B \le B_{\text{max}}$, corresponding to coarse sample structure. We typically set $B_{\text{max}}$ to 20\%-30\% of the total number of blocks $N_B$.

\item \textit{Time-based shifting:} For all DiT blocks and timesteps $t \ge t_{\text{max}}$, corresponding to high noise levels. We find $t_{\text{max}} \in [800,1000]$ to be most effective.
\end{itemize}

\noindent By tuning $B_{\text{max}}$ and $t_{\text{max}}$ on a small validation set, users can trade off between stronger control adherence and higher photorealism without any retraining. In our experiments, we prioritize realism, and perform inference-time domain shifting to improve controllability while keeping the realism loss minimal.

We observe that time-based shifting tends to have a stronger effect on realism while layer-based shifting tend to have a mild and stable effect. Therefore, to select the mentioned hyperparameters, we first fix $B_{\text{max}}$ by evaluating a small validation set across a few candidate values, ranging from 0\% to 40\% of the total blocks in 10\% increments. Then, we select $t_{\text{max}}$ to achieve the desired balance, testing values between 800 and 1000 in steps of 50. For the selection process, we evaluate realism-based metrics and define a threshold that represents the maximum reduction in realism we are willing to tolerate.
Since we do not perform an exhaustive search, the entire tuning process takes less than an hour and is performed only once.

\section{Implementation Details}
\label{app:implementation_details}
\subsection{Data}
\label{app:data_details}
\textbf{Realistic Training Data.}
As mentioned in the main paper, to create the realistic dataset, we use the same prompts from the synthetic dataset and generate $V=4$ photorealistic images using the base T2I model. This ensure fair evaluation, as both datasets contain the same diversity of objects.

As the base model is biased towards the frontal view, we generate each prompt using four viewpoint descriptions (``front'', ``back'', ``right side'', ``left side''), thus collecting $V=4$ photorealistic images per prompt.
The model often fails to follow the desired viewpoint, yet this strategy helps to enrich the data and increase pose variation.

\noindent These view descriptions are \textbf{not} used during training. During training, the prompts associated to the images do not contain this information, and the random placement of realistic images in the $2\times2$ grid is not affected by this information.

\noindent\textbf{ImageNet Curation Process for Real-World Realism Metrics}
While the prior preservation metrics indicate whether the generated images seem realistic, these metrics are based on synthesized images. We aim to ground our evaluation in real-world image data which, to the best of our knowledge, was not used for training our base model nor the pretrained models.

To that end, we processed the prompts describing the evaluation objects through LLM, to find the most fitting categories in the ImageNet dataset, resulting in 42 classes. Then, we use CLIP~\cite{radford2021learning} score to select 4 images per class from the validation set, and segment them using an internal tool similar to Rembg~\cite{rembg}.

\subsection{Training and Sampling}
\label{app:method_details}
\noindent\textbf{Training.} All evaluated models are trained for 10 epochs on 64 NVIDIA H100 GPUs, using batch size of 8 and learning rate of $5e^-5$.
Adapters that are trained separately are trained for 3 epochs.
We verified that all models and adapters converged and did not reach overfitting.
Importantly, throughout the training process we rely only on the standard reconstruction diffusion loss, which is well-established and empirically stable.
To prevent adapter-based methods from overfitting to white-background, object-centric layouts, we perform a very short warmup on realistic samples in all experiments that incorporate real data during training.

\noindent\textbf{Sampling.} For sampling we use DDIM~\cite{song2021ddim} with 50 steps.
Following the base model, we use CFG~\cite{ho2022classifier} only for the text condition.

\noindent\textbf{Hyperparameters.}
For all experiments we set $r=8$ for \adapterfull modules.

\noindent For Representation Binding, we set $\tau_B$ to be 40\% of the total number of blocks and $p_b=0.1$.

\noindent For Inference-time Domain Shifting we set $B_\text{max}$ to be 30\% of the total number of blocks and $t_\text{max} = 950$.
The selection process of $B_\text{max}$ and $t_\text{max}$ happens at test-time using a small validation set of 20 held-out objects.

\section{Evaluation}
\subsection{Implementation Details}
\label{app:eval_details}
\textbf{SDEdit Baseline.} We evaluate the effectiveness of the SDEdit~\cite{meng2021sdedit} mechanism for multiview applications. Specifically, we take the outputs of the synthetic full fine-tuning baseline, add noise to the images, and then denoise them independently using the base pretrained T2I model. The same noise realization is applied to all images of the same grid, while each image is denoised independently. Following the original paper, we set the noise injection timestep to 500. At higher timesteps, images are expected to exhibit reduced 3D consistency while potentially improving realism. However, even at $t=500$, SDEdit yields inferior performance to \method in both 3D consistency and realism.

\noindent \textbf{Pretrained Models.} For Text-to-Multiview Generation, we also evaluate TRELLIS~\cite{Xiang2024Structured3L}, a 3D-native model. We use the official text-to-3D "TRELLIS-text-large" model with the original hyperparameters. TRELLIS directly generates a 3D asset and not multiview images, therefore we generate a 3D asset from text and render the 4 orthogonal views (front, back and sides views) that we show across the paper.

\noindent \textbf{Qualitative Results.} For all baselines reported with two rank settings (32 and 128), all results shown in the main paper and the supplementary materials were obtained using rank 32.

\subsection{Ablation Study: Qualitative Results}
\begin{figure*}
\centering
\centerline{
    \includegraphics[width=0.9\linewidth]{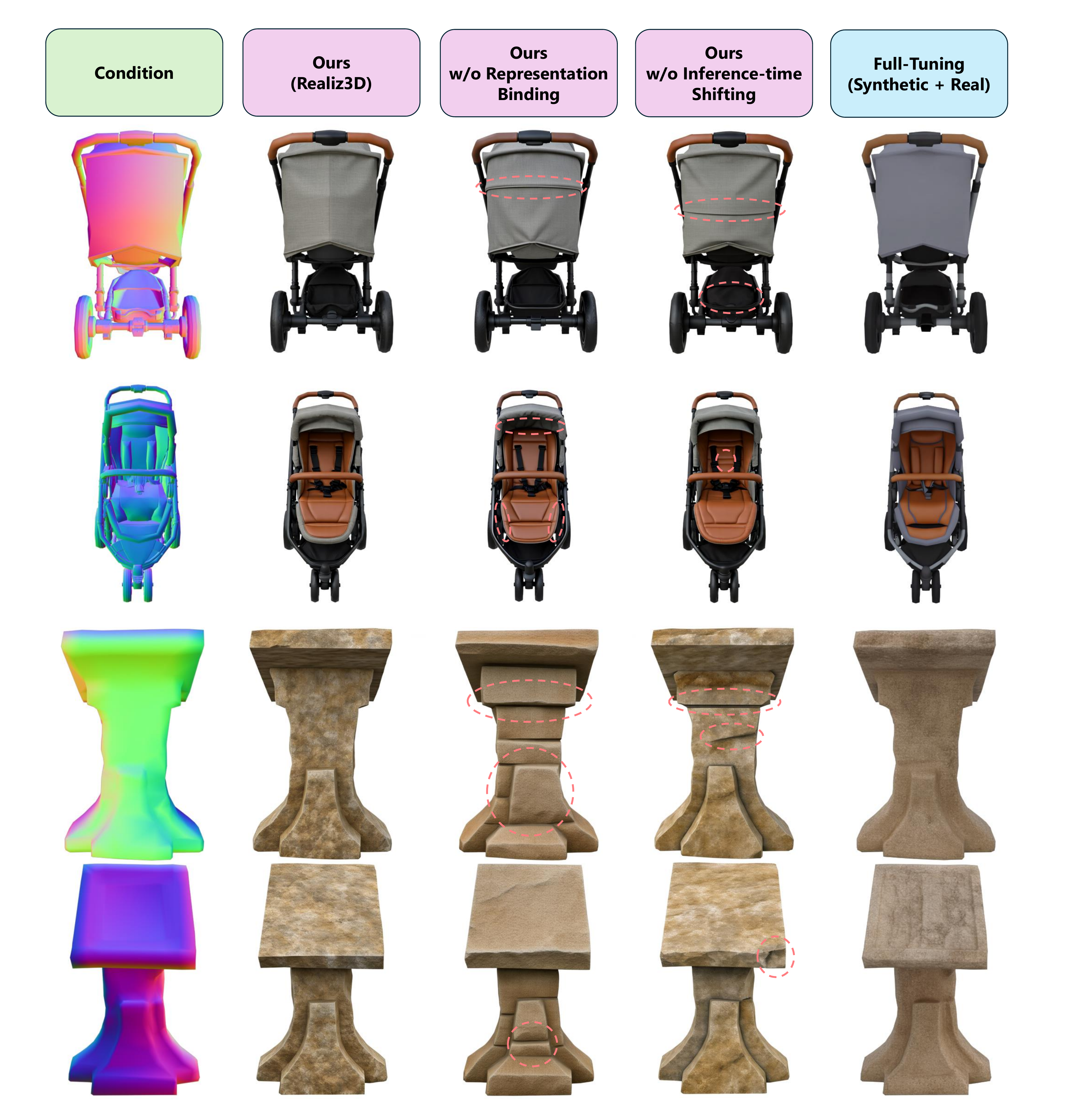}
}
\caption{\textbf{Ablation Study.} We demonstrate the importance of our Representation Binding and Inference-time Domain Shifting on Multiview Texturing. Red circles highlight inconsistent regions with the geometry. Both techniques enhance control adherence, while maintaining realism. The presented prompts: "A baby stroller with a leather seat and a black plastic container on the bottom", "A structure with a flat top, made of natural stone".}%
\label{fig:ablation}
\end{figure*}

To further support the ablation study from the main paper, we present visual examples in Fig.~\ref{fig:ablation} that illustrate the effects of our Representation Binding and Inference-Time Domain Shifting on the model’s final outputs. Both techniques improve control adherence while preserving strong realism.

\subsection{Additional Qualitative Results}
\label{app:additional_qualitative}
We present qualitative results using the evaluation data described in the main paper, as well as additional test objects from our internal dataset that were held out for evaluation and not used during training.

\begin{figure*}
\centering
\centerline{
    \includegraphics[width=0.8\linewidth]{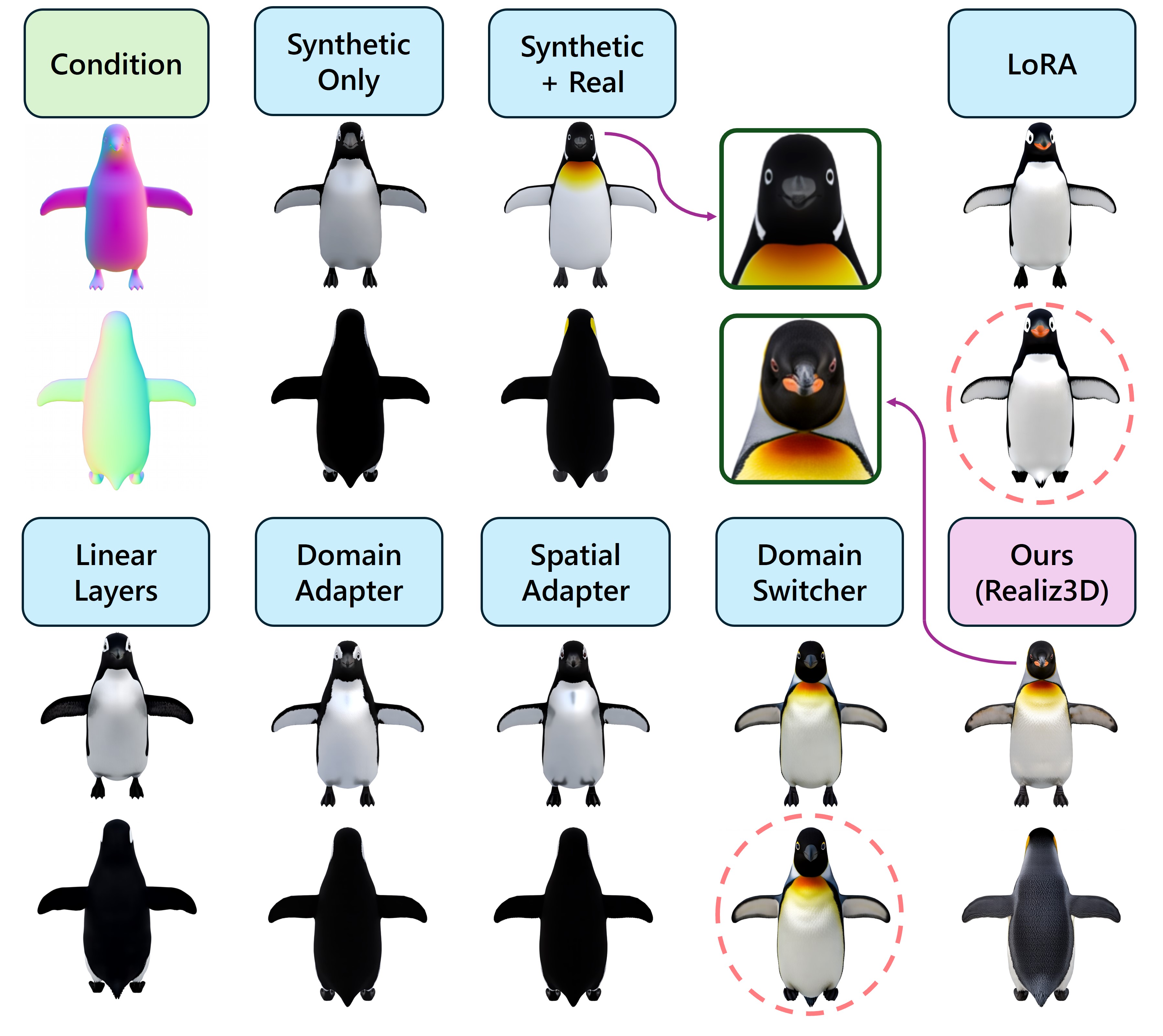}
}
\caption{\textbf{Multiview Texturing.} "A king penguin, highly realistic and detailed". Red circles highlight inconsistent regions (either with the geometry or with other views). Best viewed zoomed in.}%
\label{fig:texture_penguin}
\end{figure*}

\begin{figure*}
\centering
\centerline{
    \includegraphics[width=0.8\linewidth]{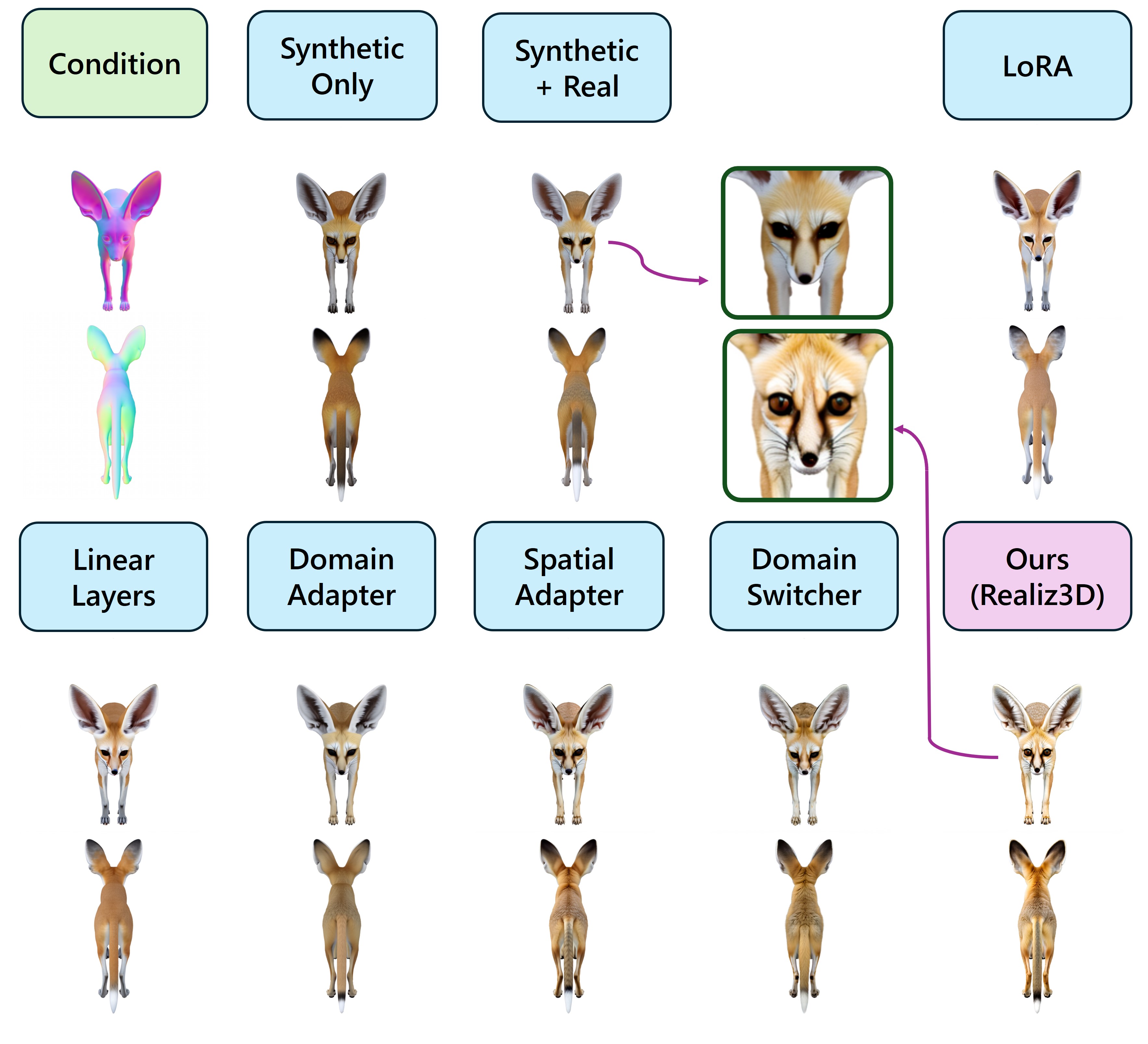}
}
\caption{\textbf{Multiview Texturing.} "A fennec fox, highly realistic and detailed". Best viewed zoomed in.}%
\label{fig:texture_fox}
\end{figure*}

\begin{figure*}
\centering
\centerline{
    \includegraphics[width=0.8\linewidth]{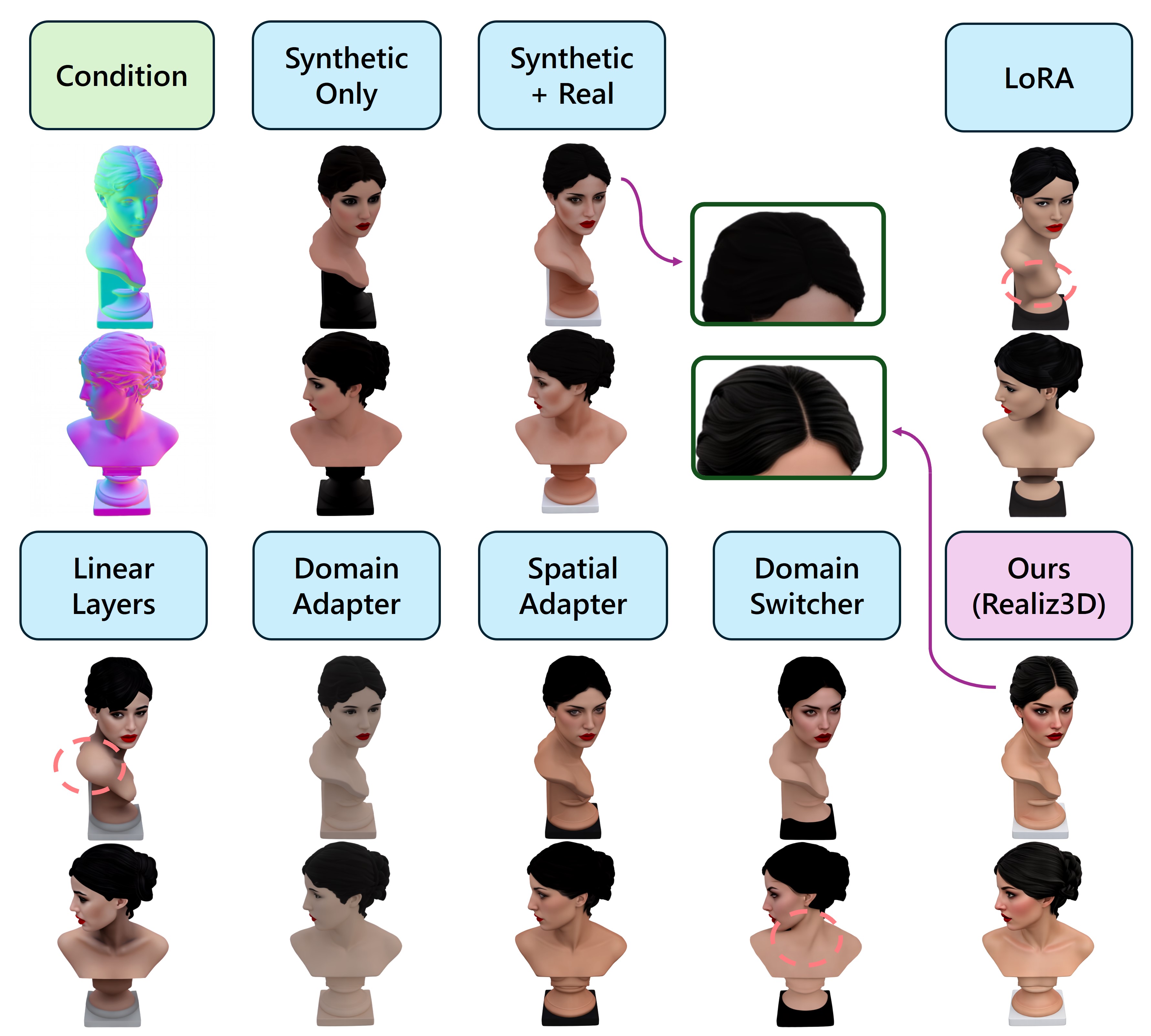}
}
\caption{\textbf{Multiview Texturing.} "A woman with a dark hair and red lips, highly realistic and detailed". Red circles highlight inconsistent regions (either with the geometry or with other views). Best viewed zoomed in.}
\label{fig:texture_woman}
\end{figure*}

\noindent\textbf{Multiview Texturing.} We present additional qualitative results in~\cref{fig:texture_penguin}, \cref{fig:texture_fox} and \cref{fig:texture_woman}, showing all baselines described in the main paper. The corresponding prompt appears in the caption. Although our method uses both normal and position maps as geometric conditions, only the normal maps are shown for readability. \method achieves significant improvements in photorealism while remaining 3D-consistent and faithful to the geometric conditions.

\begin{figure*}
\centering
\centerline{
    \includegraphics[width=\linewidth]{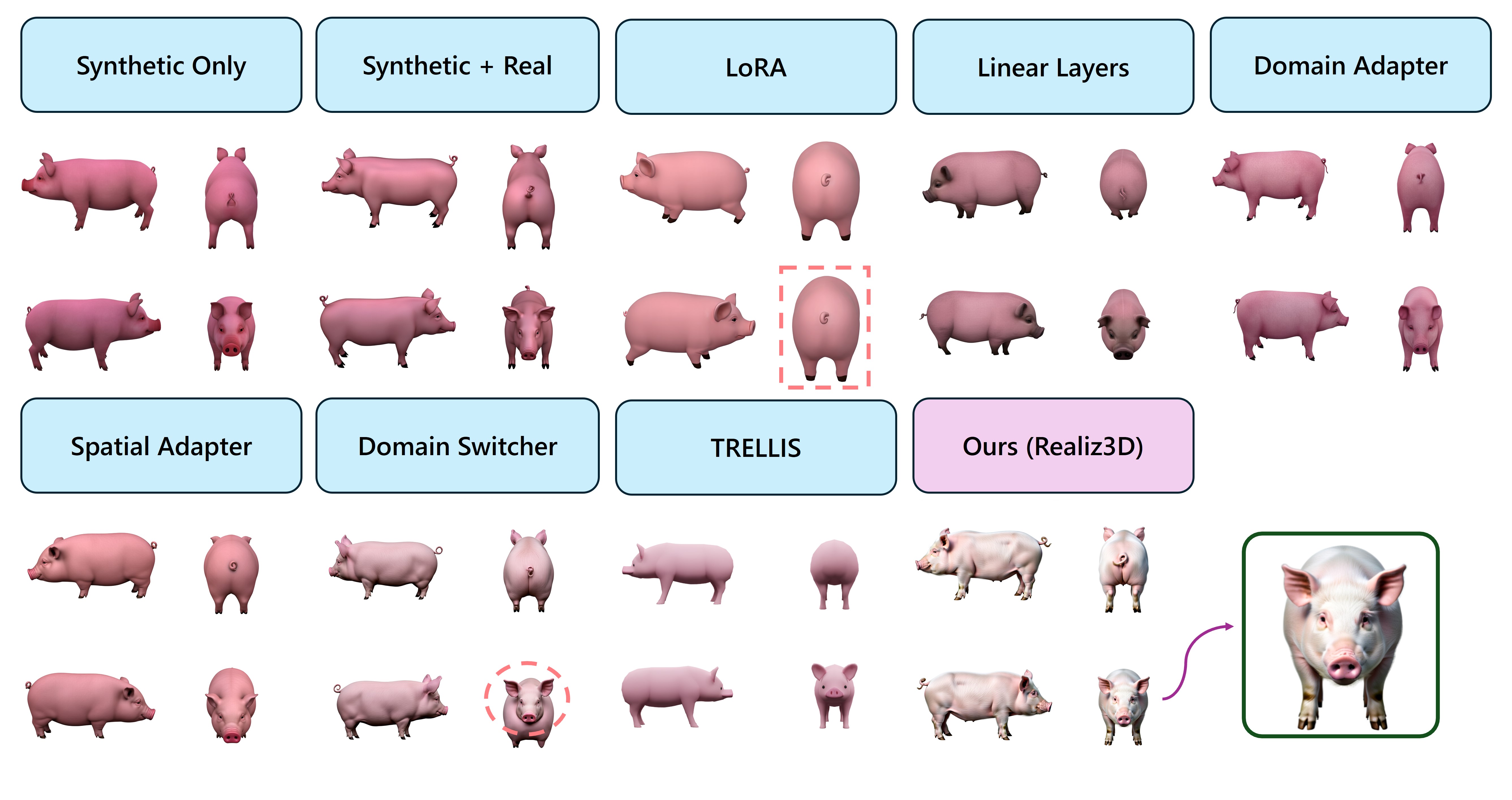}
}
\caption{\textbf{Text-to-Multiview Generation.} "A pink farm pig, highly realistic and detailed". Red circles/squares highlight inconsistent regions/incorrect viewpoints, respectively. Best viewed zoomed in.}
\label{fig:t2mv_pig}
\end{figure*}

\begin{figure*}
\centering
\centerline{
    \includegraphics[width=\linewidth]{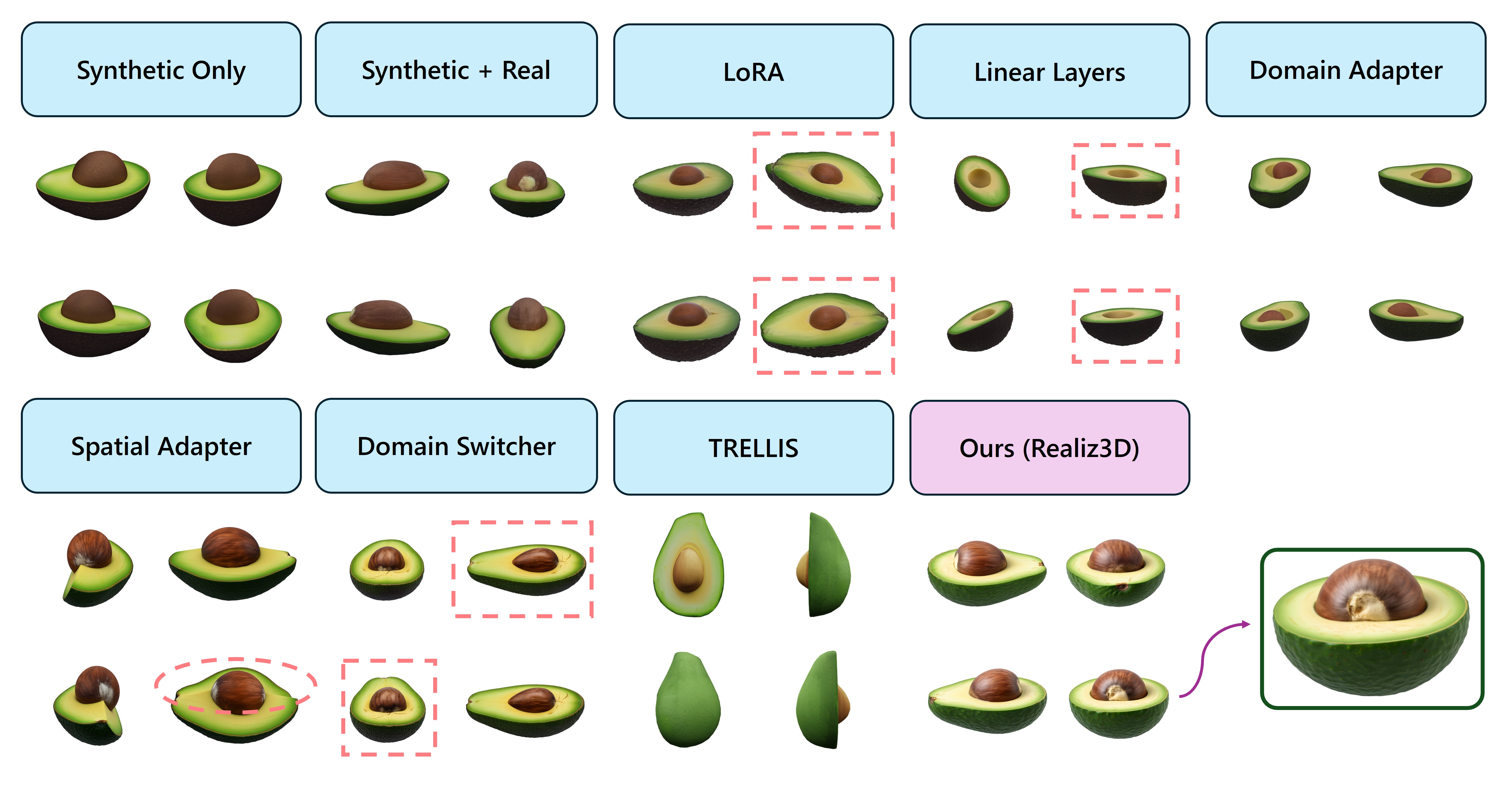}
}
\caption{\textbf{Text-to-Multiview Generation.} "A half slice of avocado, highly realistic and detailed". Red circles/squares highlight inconsistent regions/incorrect viewpoints, respectively. Best viewed zoomed in.}%
\label{fig:t2mv_avocado}
\end{figure*}

\begin{figure*}
\centering
\centerline{
    \includegraphics[width=\linewidth]{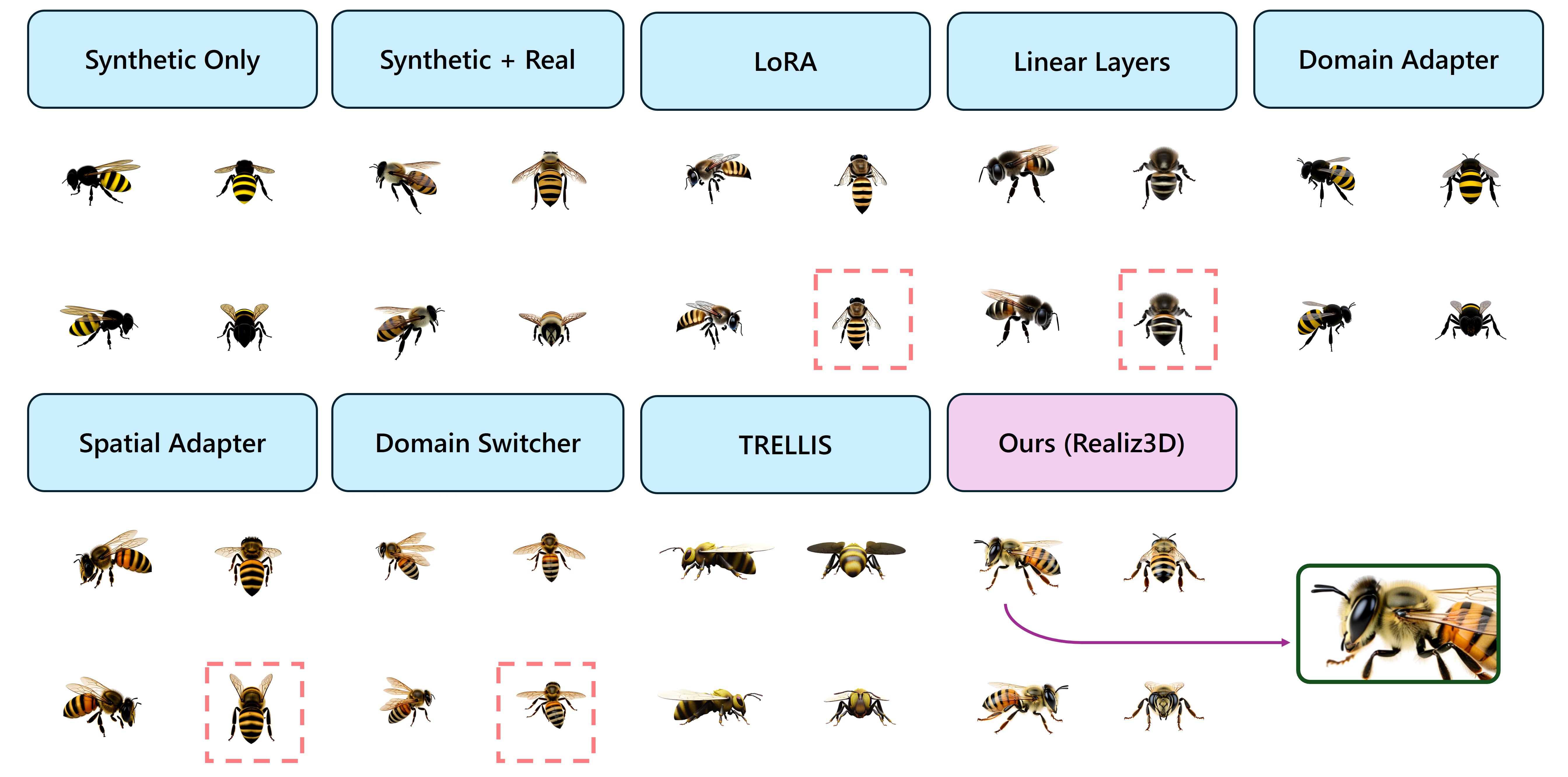}
}
\caption{\textbf{Text-to-Multiview Generation.} "A bee, highly realistic and detailed". Red circles/squares highlight inconsistent regions/incorrect viewpoints, respectively. Best viewed zoomed in.}
\label{fig:t2mv_bee}
\end{figure*}

\noindent\textbf{Text-to-Multiview Generation.} We present additional qualitative results in~\cref{fig:t2mv_pig}, \cref{fig:t2mv_avocado} and \cref{fig:t2mv_bee}, showing all baselines described in the main paper. The corresponding prompt appears in the caption. \method achieves notable improvements in photorealism while maintaining strong 3D consistency.

\subsection{Text-to-3D Results}
We perform text-to-3D generation by backprojecting generated textures onto their corresponding original meshes. Note that we generate only four orthogonal views, which may not fully cover the entire surface.

\textbf{The results are provided on our project page}, where we present side-by-side 3D assets produced by the full-tuning baseline (trained on both real and synthetic data for fairness) and by \method.

\noindent The results highlight that \method achieves comparable 3D-consistency to the fully synthetic and full fine-tuning baselines, producing coherent and realistic 3D assets.

\section{Limitations and Future Work}
\label{app:limitations}
\begin{figure}[H]
\centering
\centerline{
    \includegraphics[width=\linewidth]{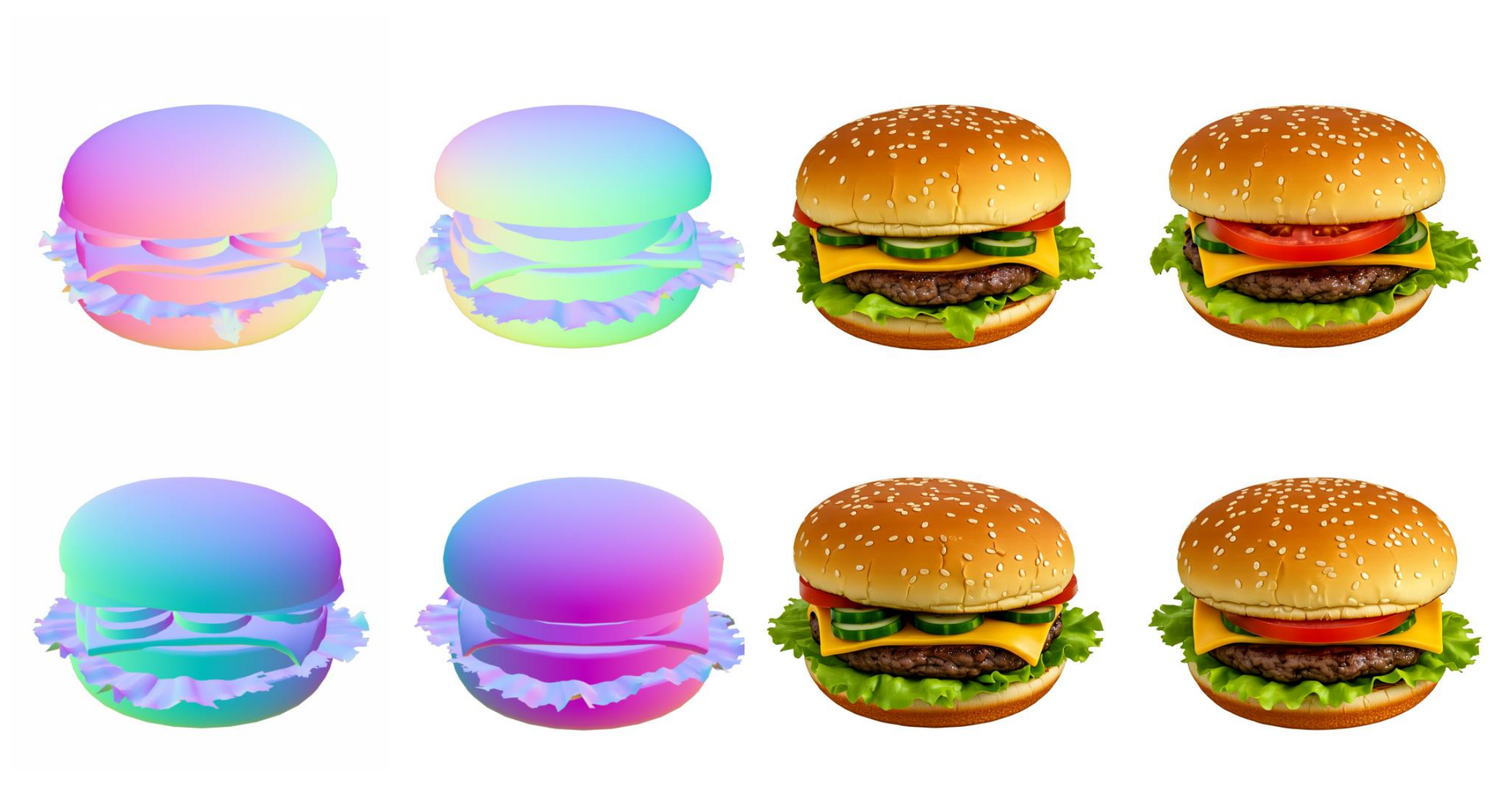}
}
\caption{\textbf{Multiview Texturing.} Inconsistent lighting caused by the base T2I model's lighting bias.}
\label{fig:failure_case}
\end{figure}

As mentioned in the main paper, although \method significantly improves realism, a small gap in control adherence remains. We attribute this to several key factors:
(1) 3D consistency is sensitive to fine-grained details. Because the synthetic data primarily contains smooth textures, the task becomes easier for the model. As a result, synthetic baselines tends to produce relatively smooth outputs, reducing the need to learn perfect pixel-level 3D consistency. In other words, these baselines often appear consistent even without having learned accurate pixel-level 3D-consistency. In contrast, \method produces fine-grained details (e.g., complex textures and materials, hair and fur) that are much more sensitive to even slight misalignment.
(2) Domain gaps can occasionally manifest in unrealistic geometry, not just appearance, causing \method to slightly deviate from the original geometry.
(3) The base model’s lighting bias can lead to inconsistent appearance. Our synthetic data is uniformly lit, and the generated views largely preserve this property, appearing evenly illuminated. However, the base T2I model shows a bias in lighting for certain objects and materials. An example failure case appears in~\cref{fig:failure_case}, where we generate the texture of a hamburger. The hamburger is consistently lit more strongly from the front right, while the back left remains noticeably darker. The same trend is consistent across different seeds. After examining hamburger images online and inspecting those generated by our base T2I model, we find that this lighting condition is extremely common in real-world images, creating a strong lighting bias in the base model.
Recent advances in relighting~\cite{liang2025diffusion, chaturvedi2025synthlight, litman2025lightswitch} and specifically the use of two synthetic domains, one uniformly lit and one randomly lit, offer promising avenues for addressing this gap.

In addition, a natural extension of \method is to apply our techniques to video diffusion models, which have recently demonstrated remarkable capabilities. When these models are trained to incorporate a 3D condition, they are often fine-tuned on synthetic data, introducing a similar domain gap.

\end{document}